\documentclass[pra,aps,twocolumn,amsfonts,showpacs,superscriptaddress]{revtex4-1}

\usepackage{epsfig} 
\usepackage{psfrag} 
\usepackage{amsmath}
\usepackage{amssymb} 
\usepackage{color} 
\usepackage{graphicx}
\usepackage{hyperref} 
\usepackage{subfigure}

\begin{document}

\title{Quantum Phase Transitions in Bosonic Heteronuclear Pairing
  Hamiltonians}

\author{M. Hohenadler} 
\affiliation{Institute for Theoretical Physics and Astrophysics, 
University of W{\"u}rzburg, Germany.}
\author{A. O. Silver} 
\affiliation{University of Cambridge, Cavendish Laboratory, 
Cambridge, CB3 0HE, UK.}
\author{M. J. Bhaseen} 
\affiliation{University of Cambridge, Cavendish Laboratory, 
Cambridge, CB3 0HE, UK.}
\author{B. D. Simons} 
\affiliation{University of Cambridge, Cavendish Laboratory, 
Cambridge, CB3 0HE, UK.}

\date{\today}

\begin{abstract}
  We explore the phase diagram of two-component bosons with Feshbach
  resonant pairing interactions in an optical lattice.  It has been
  shown in previous work to exhibit a rich variety of phases and phase
  transitions, including a paradigmatic Ising quantum phase transition
  within the second Mott lobe.  We discuss the evolution of the phase
  diagram with system parameters and relate this to the predictions of
  Landau theory.  We extend our exact diagonalization studies of the
  one-dimensional bosonic Hamiltonian and confirm additional Ising
  critical exponents for the longitudinal and transverse magnetic
  susceptibilities within the second Mott lobe.  The numerical results
  for the ground state energy and transverse magnetization are in good
  agreement with exact solutions of the Ising model in the
  thermodynamic limit. We also provide details of the low-energy
  spectrum, as well as density fluctuations and superfluid fractions
  in the grand canonical ensemble.
\end{abstract}


\maketitle

\section{Introduction}

In the last few years there has been considerable experimental and
theoretical interest in studying Feshbach resonances between different
atomic species and isotopes. This activity encompasses pairing
interactions and molecule formation in a wide variety of Fermi--Fermi,
Bose--Fermi, and Bose--Bose mixtures. Such systems provide many
possibilities for novel phases and phenomena, ranging from dipolar
condensates \cite{Santos:Dipolar,Baranov:Dipolar} to highly
controllable chemical reactions \cite{Ospelkaus:Reactions}.  Recent
examples include heteronuclear resonances in $^{85}{\rm
  Rb}$--$^{87}{\rm Rb}$
\cite{Papp:Hetero,Papp:Tunable,Ticknor:Multiplet}, $^{41}{\rm
  K}$--$^{87}{\rm Rb}$
\cite{Weber:Assoc,Thalhammer:Double,Thalhammer:Coll,Barontini:Obs,Catani:Entropy,Lamporesi:Mixed},
$^{39}{\rm K}$--$^{87}{\rm Rb}$ \cite{Simoni:Near}, and $^{87}{\rm
  Rb}$--$^{133}{\rm Cs}$ \cite{Pilch:Obs} Bose mixtures over a range
of experimental parameters.

Motivated by these developments, we recently investigated the phase
diagram of two-component bosons pairing in an optical lattice
\cite{Bhaseen:Feshising}. Amongst our findings, we identified a
paradigmatic Ising quantum phase transition occurring within the
second Mott lobe. The principal aim of this manuscript is to provide a
more detailed overview of this heteronuclear system, and to extend the
scope of physical observables presented in
Ref.~\cite{Bhaseen:Feshising}. We expand our previous exact
diagonalization results in several directions, and confirm the
additional Ising critical exponents, $\alpha=0$, $\gamma=7/4$ and
$\delta=15$. We also provide results for the low-energy spectrum, as
well as density fluctuations and superfluid fractions in the
grand-canonical ensemble. We supplement this with a discussion of the
evolution of the phase diagram with system parameters, and its direct
connection to Landau theory.

The layout of this manuscript is as follows. In Sec.~\ref{Sect:Model}
we describe the heteronuclear model with Feshbach interactions.  In
Secs.~\ref{Sect:PD} and \ref{Sect:Landau} we discuss the mean field
phase diagram and the Landau theory description.  In
Sec.~\ref{Sect:Magnetic} we focus on the Mott states and present a
derivation of the effective quantum Ising model.  We confirm these
findings in Sec.~\ref{Sect:Num} by exact diagonalization of the
one-dimensional bosonic Hamiltonian.  We conclude in
Sec.~\ref{Sect:Conc}. We incorporate the principal results of
Ref.~\cite{Bhaseen:Feshising}.

\section{The Model}
\label{Sect:Model}
We consider two-component bosons with a ``spin'' index
$\downarrow,\uparrow$, which may be different hyperfine states,
isotopes or species. These components may form molecules, $m$, as
described by the Hamiltonian
\begin{equation}
\begin{aligned}
H & =\sum_{i\alpha}\epsilon_\alpha n_{i\alpha}- \sum_{\langle
ij\rangle} \sum_\alpha t_\alpha\left(a_{i\alpha}^\dagger
a_{j\alpha}+{\rm h.c.} \right)\\ & \hspace{1cm}
+\sum_{i\alpha\alpha^\prime}
\frac{U_{\alpha\alpha^\prime}}{2}:n_{i\alpha}n_{i\alpha^\prime}: +
\,H_\text{F}.
\label{atmolham}
\end{aligned}
\end{equation}
Here, $a_{i\alpha}$ are Bose annihilation operators, where $i$ labels
the lattice sites, $n_{i\alpha}=a_{i\alpha}^\dagger a_{i\alpha}$, and
$\alpha=\downarrow,\uparrow,m$; $\epsilon_\alpha$ are on-site
potentials, $t_\alpha$ are hopping parameters, $\langle ij\rangle$
denotes summation over nearest neighbor bonds, and
$U_{\alpha\alpha^\prime}$ are interactions. Molecule formation is
described by the s-wave interspecies Feshbach resonance term
\begin{equation}
H_\text{F}= g\sum_i(a^\dagger_{im} a_{i\uparrow}a_{i\downarrow}+{\rm h.c.}).
\label{HF}
\end{equation}
Similar problems have been studied in the continuum limit with s-wave
\cite{Zhou:PS} and p-wave resonances \cite{Radzi:spinor}.  Closely
related homonuclear systems have also been considered in the continuum
\cite{Timmermans:Feshbach,Rad:Atmol,Romans:QPT,Koetsier:Bosecross,Radzi:Resonant}
and on the lattice
\cite{Dickerscheid:Feshbach,Sengupta:Feshbach,Rousseau:Fesh,Rousseau:Mixtures}.
Normal ordering implies ${:n_{i\alpha}n_{i\alpha}:} =
n_{i\alpha}(n_{i\alpha}-1)$ for like species, and
${:n_{i\alpha}n_{i\alpha^\prime}:} = n_{i\alpha}n_{i\alpha^\prime}$
for distinct species. For simplicity we consider hardcore atoms and
molecules and set $U_{m\uparrow}=U_{m\downarrow}\equiv U$ and
$U_{\uparrow\downarrow}\equiv V$. We begin work in the grand canonical
ensemble with $H_\mu=H-\mu_{\rm T} N_{\rm T}-\mu_{\rm D}N_{\rm D}$,
where $N_{\rm T}\equiv \sum_i
(n_{i{\uparrow}}+n_{i\downarrow}+2n_{im})$ is the total atom number,
including a factor of two for molecules, and $N_{\rm D}\equiv
\sum_i(n_{i\uparrow}-n_{i\downarrow})$ is the up-down population
imbalance.  The chemical potentials may be absorbed into the
coefficients, $\tilde\epsilon_\uparrow\equiv\epsilon_\uparrow-\mu_{\rm
  T}-\mu_{\rm D}$,
$\tilde\epsilon_\downarrow\equiv\epsilon_\downarrow-\mu_{\rm
  T}+\mu_{\rm D}$, and $\tilde\epsilon_m\equiv\epsilon_m-2\mu_{\rm
  T}$.  For the numerical investigation of the Ising transition we
shall subsequently switch to the canonical ensemble with a fixed total
density, $\rho_{\rm T}\equiv n_\uparrow+n_\downarrow+2n_m=2$.

\section{Phase Diagram}
\label{Sect:PD}
\begin{figure} 
  \includegraphics[width=3in]{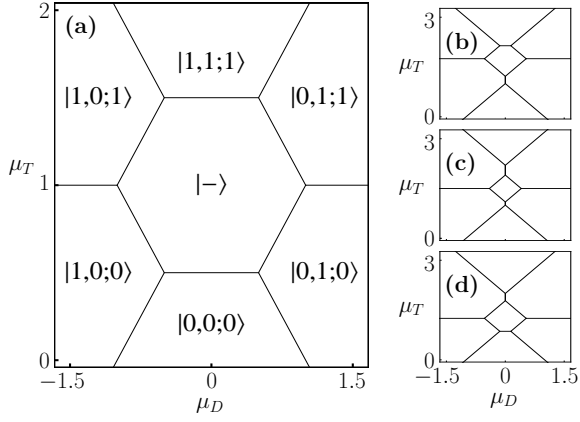}
  \caption{Zero hopping phase diagram showing the minimum energy
    eigenstates in the basis
    $|n_\downarrow,n_\uparrow;n_m\rangle$. The Feshbach coupling mixes
    $|1,1;0\rangle$ and $|0,0;1\rangle$ to yield $|\pm\rangle$ with
    energies $E_\pm=\tilde\epsilon_m-h/2\pm\sqrt{(h/2)^2+g^2}$, where
    $h\equiv \epsilon_m-\epsilon_\downarrow-\epsilon_\uparrow-V$. The
    other states have
    $E(n_\downarrow,n_\uparrow;n_m)=\sum_\alpha\tilde\epsilon_\alpha
    n_\alpha+Vn_\uparrow
    n_\downarrow+Un_m(n_\uparrow+n_\downarrow)$. The total density
    $\rho_{\rm T}\equiv n_\downarrow+n_\uparrow+2n_m$ is pinned to
    integer values and increases with $\mu_{\rm T}$. Increasing
    (decreasing) $\mu_{\rm D}$ favors up (down) atoms. The topology
    changes with the system parameters and depends on the signs of the
    energy gaps, $\Delta_\pm$, defined in the text. We set
    $\epsilon_\downarrow=\epsilon_\uparrow=g=1$, $U=0$ and (a)
    $\epsilon_m=2$, $V=0$, (b) $\epsilon_m=3.5$, $V=1$, (c)
    $\epsilon_m=3$, $V=1.2$, (d) $\epsilon_m=2.5$ $V=1$. The panels
    shown in Fig.~\ref{Fig:MFT} correspond to vertical slices through diagram 
(a).}
\label{Fig:Zerohop}
\end{figure}
\begin{figure}
\begin{center}
  \includegraphics[clip,width=3in]{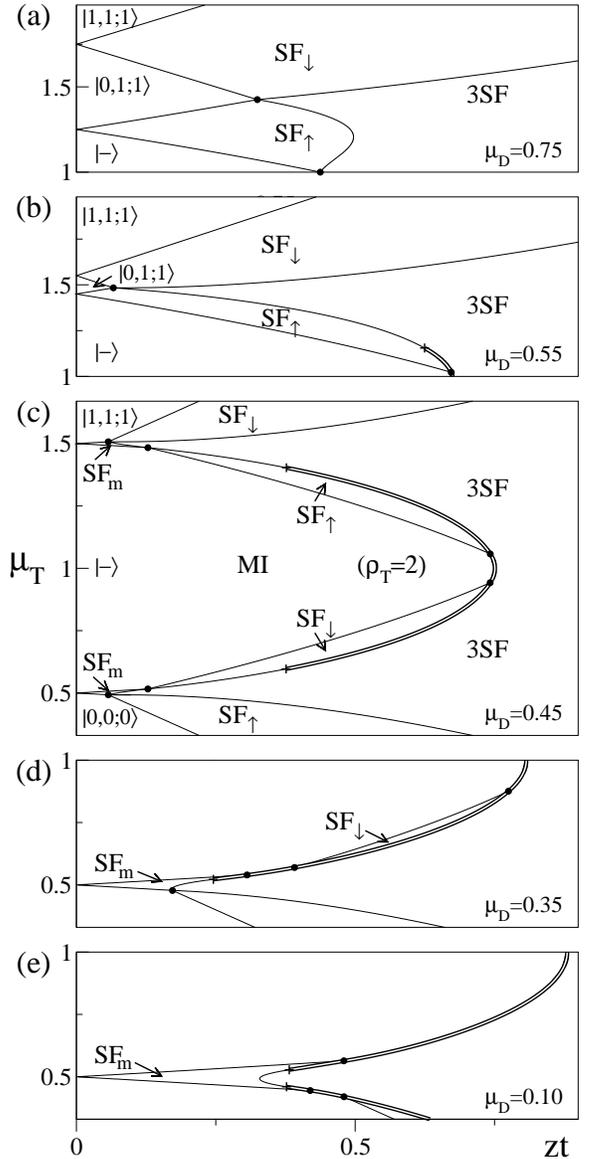}
\end{center}
\caption{Evolution of the mean field phase diagram with $\mu_{\rm D}$,
  corresponding to vertical scans through Fig.~\ref{Fig:Zerohop}(a).
  We set $\epsilon_\uparrow=\epsilon_\downarrow=1$, $\epsilon_m=2$,
  $g=1$, $U=V=0$, $t_\uparrow=t_\downarrow=t$, and $t_m=t/2$. We
  indicate the one-component up, down and molecular superfluids, by
  ${\rm SF}_\uparrow$, ${\rm SF}_\downarrow$ and ${\rm SF}_m$, while
  MI denotes a Mott insulator.  Phase 3SF has all three components
  superfluid. We denote first order (continuous) transitions by double
  (single) lines. Junctions between phases are indicated by a dot, and
  the termination of first order lines by a cross.}
\label{Fig:MFT}
\end{figure}
\begin{figure}
\includegraphics[width=3in]{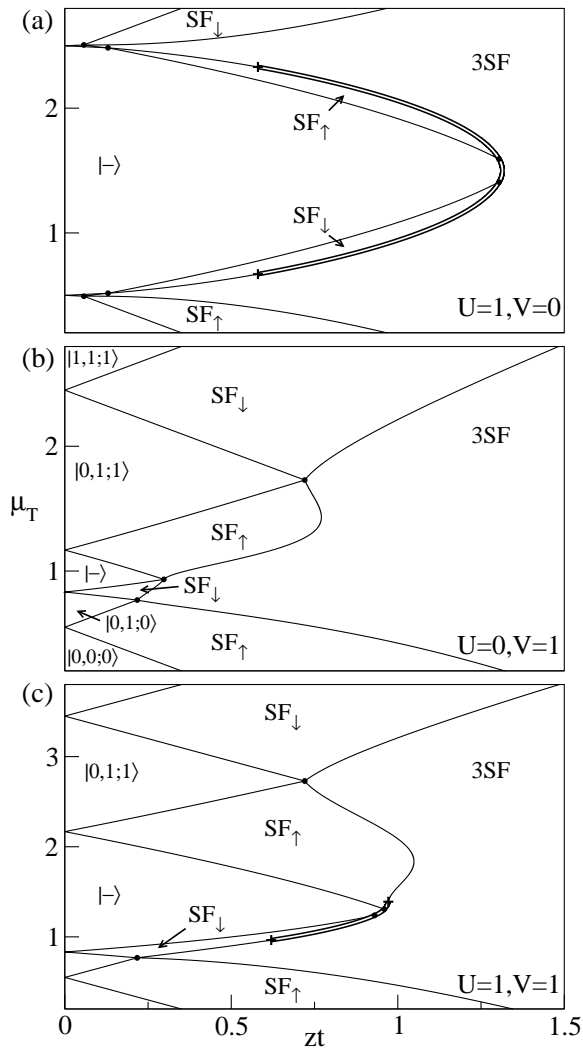}
\caption{Mean field phase diagram in the presence of interactions, $U$
  and $V$.  We set $\epsilon_\uparrow=\epsilon_\downarrow=1$,
  $\epsilon_m=2$, $g=1$, $t_\uparrow=t_\downarrow=t$, $t_m=t/2$ and
  $\mu_{\rm D}=0.45$ as in Fig.~\ref{Fig:MFT}(c), and consider the
  effects of $U$ and $V$ in turn. (a) $U=1$, $V=0$, showing the
  increase in extent of the $|-\rangle$ Mott lobe. (b) $U=0$, $V=1$,
  showing the appearance of asymmetry. (c) $U=1$, $V=1$ showing the
  combined effect of both interactions. The characteristic phases and phase transitions mirror those seen in the $U=V=0$ limit shown in Fig.~\ref{Fig:MFT}.}
\label{Fig:MFTUV}
\end{figure}
As discussed in Ref.~\cite{Bhaseen:Feshising}, in elucidating the zero
temperature phase diagram the zero hopping limit provides a useful
anchor point.  In particular, the topology of the zero hopping phase
diagram depends on the Hamiltonian parameters via the energy difference between
proximate phases; see Fig.~\ref{Fig:Zerohop}.  In this respect it is
convenient to define the energy gap, $\Delta_-\equiv E(0,0;0)-E_-$,
between the vacuum state and the second Mott lobe, evaluated at the
intersection point $E(1,0;0)=E(0,1;0)=E(0,0;0)$, where we label the
energies $E(n_\downarrow,n_\uparrow;n_m)$ in the occupation
basis. Similarly, we define $\Delta_+\equiv E_--E(1,1;1)$, between the
second and upper Mott lobes, where $E(1,0;1)=E(0,1;1)=E_-$. This
yields $\Delta_\pm=\sqrt{g^2+(h/2)^2}-V\pm h/2$, where $h\equiv
\epsilon_m-\epsilon_\downarrow-\epsilon_\uparrow-V$.  One obtains the
structure shown in Fig.~\ref{Fig:Zerohop}(a) for $\Delta_\pm>0$, (b)
for $\Delta_+>0$ and $\Delta_-<0$, (c) for $\Delta_\pm<0$, and (d) for
$\Delta_+<0$ and $\Delta_->0$.

For simplicity, we begin with the parameters used in
Fig.~\ref{Fig:Zerohop}(a), where $U=V=0$. Since we include Feshbach
resonant interactions, $g$, this limit captures many of the principal
features and phases of the interacting problem, including the presence
of Mott states. We will incorporate the effects of finite $U$ and $V$
in the subsequent discussion. The mean field phase
diagram is obtained by minimizing the effective Hamiltonian
\begin{equation}
H=H_0-\sum_\alpha zt_\alpha\left(a_\alpha^\dagger\phi_\alpha 
+\phi_\alpha^\ast a_\alpha-
|\phi_\alpha|^2\right),
\label{MFTham}
\end{equation}
where $H_0$ is the single site zero hopping contribution to
(\ref{atmolham}), $z$ is the coordination, and $\phi_\alpha\equiv
\langle a_{i\alpha}\rangle$; see Fig.~\ref{Fig:MFT}.  The phase
diagram is symmetric under $\mu_{\rm T}\rightarrow 2-\mu_{\rm T}$ due
to invariance of the Hamiltonian (\ref{atmolham}) under particle-hole
and spin flip operations, $a_\alpha\leftrightarrow a_\alpha^\dagger$,
$\mu_{\rm T}\rightarrow \epsilon_m+U-\mu_{\rm T}$,
$a_\downarrow\leftrightarrow a_\uparrow$, when
$t_\downarrow=t_\uparrow$ and $h=0$; this extends the ``top-bottom''
symmetry in Fig.~\ref{Fig:Zerohop}(a). Likewise, the system is
invariant under $\mu_{\rm D}\rightarrow
\epsilon_\uparrow-\epsilon_\downarrow-\mu_{\rm D}$,
$a_\downarrow\leftrightarrow a_\uparrow$ and
$t_\downarrow\leftrightarrow t_\uparrow$, which extends the
``left-right'' symmetry in Fig.~\ref{Fig:Zerohop}.  The phase diagram
has a rich structure and exhibits distinct Mott insulators, single
component atomic and molecular condensates, and a phase with all three
species superfluid.  However, two-component superfluids are absent due
to the structure of the Feshbach term
\cite{Bhaseen:Feshising,Zhou:PS}.  The main panel shown in
Fig.~\ref{Fig:MFT}(c) also displays an intricate network of continuous
quantum phase transitions, quantum critical points and first order
phase transitions, where the latter are inferred by discontinuities in
the order parameters and derivatives of the ground state energy. In
particular, the first order segments shroud the second Mott lobe and
overextend beyond the junctions of the proximate superfluid phases. 
Similar features also emerge in the two-component Bose--Hubbard model
in the absence of Feshbach interactions \cite{Kuklov:Commens}.  As may
be seen by tracking the evolution with $\mu_{\rm D}$, these first
order segments emerge from an underlying tetracritical point, as shown
in panels (a) and (b). In a similar way, the tetracritical points
shown in Fig.~\ref{Fig:MFT}(c) may bifurcate into segments connected
by first order transitions as shown in panels (d) and (e).  As we
shall discuss in Sec. \ref{Sect:Landau}, the presence and
transmutation of these elementary critical points and first order
segments may be seen from a reduced two-component Landau theory.
Before embarking on this discussion, let us note that these principal
features also emerge for non-vanishing $U$ and $V$, as shown in
Fig.~\ref{Fig:MFTUV}. Although finite interactions induce quantitative
distortions of the phase diagram depicted in Fig.~\ref{Fig:MFT}, the
characteristic phases and phase transitions are nonetheless present in
the $U=V=0$ limit.
\section{Landau Theory}
\begin{figure}
\begin{center}
\includegraphics[clip,width=3in]{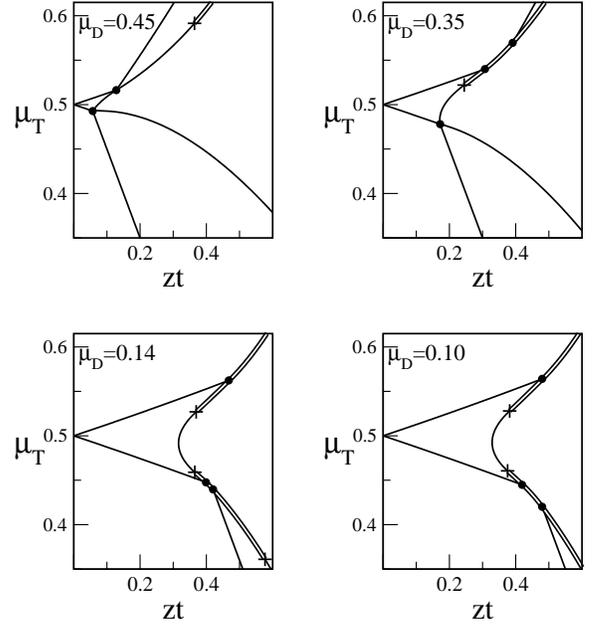}
\end{center}
\caption{Magnified portion of the lower left region of
  Fig.~\ref{Fig:MFT} showing the underlying tetracritical points as
  $\mu_{\rm D}$ is varied. The lower and middle tetracritical points
  bifurcate at $\mu_{\rm D}\approx 0.203$ and $\mu_{\rm D}\approx
  0.445$ respectively. The former corresponds to the reduced Landau
  theory criterion $\Lambda_{\uparrow
    m}=-\sqrt{\Lambda_{\uparrow\uparrow}\Lambda_{mm}}$ obtained by
  saddle point elimination of $\phi_\downarrow$ in the perturbation expansion 
around the vacuum state, $|0,0;0\rangle$. The latter corresponds to 
$\Lambda_{\downarrow m}=-\sqrt{\Lambda_{\downarrow\downarrow}\Lambda_{mm}}$ 
after eliminating $\phi_\uparrow$ in the perturbation expansion 
around the second Mott lobe, $|-\rangle$.}
\label{Fig:Tip2}
\end{figure}

\begin{figure}
\begin{center}
\includegraphics[clip,width=3in]{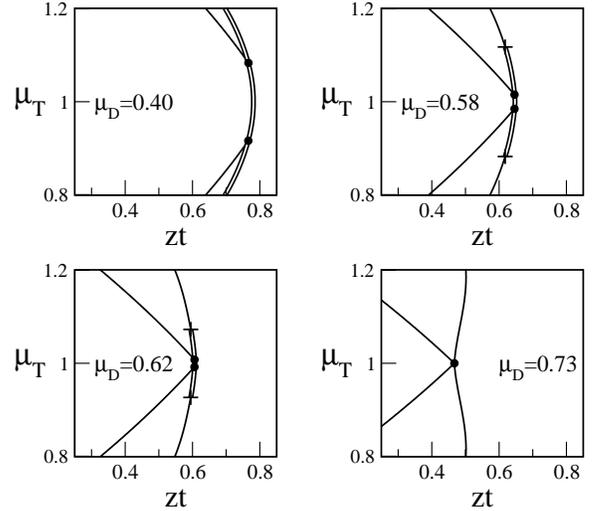}
\end{center}
\caption{Magnified portion of the second lobe tip in
  Fig.~\ref{Fig:MFT} showing the retreat of the first order
  transitions and the emergence of a tetracritical point at $\mu_{\rm
    D}\approx 0.715$. This is given by the reduced Landau theory
  criterion, $\Lambda_{\uparrow
    \downarrow}=-\sqrt{\Lambda_{\uparrow\uparrow}\Lambda_{\downarrow\downarrow}}$,
  obtained after saddle point elimination of $\phi_m$ in perturbation
  theory around the second Mott lobe, $|-\rangle$.}
\label{Fig:Tip}
\end{figure}

To understand the structure of the mean field phase diagram we develop
a Landau theory description.  The eigenenergies of the effective
single site Hamiltonian (\ref{MFTham}) may be calculated by treating
the off-diagonal hopping contribution as a perturbation,
\begin{equation}
{\mathcal V}=-\sum_\alpha zt_\alpha(a^\dagger_\alpha\phi_\alpha+\text{h.c.}),
\end{equation}
where $\alpha=\downarrow,\uparrow,m$. In principle, this may be
performed to arbitrary order using the general formalism in
Ref.~\cite{Messiah:QM}.  To illustrate the observed topology it is
sufficient to obtain the appropriate eigenenergy to fourth order
\begin{equation}
\begin{split}
  &E_r=E^0_r+\langle r|{\mathcal V} S^1_r {\mathcal V}|r\rangle+\langle r|{\mathcal V} S^1_r{\mathcal V}S^1_r {\mathcal V}|r\rangle\\
  &+\langle r|{\mathcal V}S^1_r{\mathcal V}S^1_r{\mathcal V}S^1_r
  {\mathcal V}|r\rangle -\langle r|{\mathcal V}S^1_r {\mathcal
    V}|r\rangle\langle r|{\mathcal V} S^2_r{\mathcal
    V}|r\rangle+\dots,
\label{Enexp}
\end{split}
\end{equation}
where $r$ labels the zero hopping Mott state, and the operator
$S^k_r=\sum_{n\neq r}|n\rangle\langle n|/(E^{0}_r-E^{0}_n)^k$ is
introduced \cite{Messiah:QM}. The expansion (\ref{Enexp}) takes on a
simplified form due to the off-diagonal nature of the
perturbation. One obtains
\begin{equation}
\begin{split}
  E=E_0&+\frac{1}{2}\sum_\alpha m_\alpha |\phi_\alpha|^2+\frac{\gamma}{2}\left(\phi^*_m\phi_\uparrow\phi_\downarrow+\text{h.c.}\right)\\
  &+\frac{1}{4}\sum_{\alpha\beta}\lambda_{\alpha\beta}|\phi_\alpha|^2|\phi_\beta|^2+\mathcal{O}(\phi^6),
\label{Eexp}
\end{split}
\end{equation}
where the explicit parameters, but not the overall structure, depend
on the unperturbed Mott state. The transition to the one-component
superfluids is determined by the vanishing of the quadratic mass
terms; expressions for these coefficients are given in
App. \ref{App:Landau}.  The function (\ref{Eexp}) is minimized when
$\arg(\gamma\phi^*_m\phi_\downarrow\phi_\uparrow)=\pi$, and without loss of
generality we may take the fields to be real.  To see how a Landau
theory of this type gives rise to the observed evolution of the
tetracritical points we consider the situation where one of the fields
is massive with $m\gg 0$.  Replacing this field by its saddle point
solution one obtains the energy in terms of the two remaining
fields. As a concrete example let us consider the vicinity of the
lower tetracritical point shown in Fig.~\ref{Fig:MFT}(c), where
$m_\downarrow \gg 0$. The value of $\phi_\downarrow$ which minimizes
the energy is given by
\begin{equation}
\begin{split}
  0=\frac{\partial E}{\partial
    \phi_\downarrow}&=m_\downarrow\phi_\downarrow
  +\gamma \phi_m \phi_\uparrow +\lambda_{\downarrow\downarrow}\phi_\downarrow^3\\
  &+\lambda_{\downarrow\uparrow}\phi_\downarrow\phi_\uparrow^2+\lambda_{\downarrow
    m}\phi_\downarrow \phi_m^2.
\end{split}
\end{equation}
Taking $\phi_\downarrow=-\gamma \phi_m
\phi_\uparrow/m_\downarrow+\mathcal{O}(\phi^4)$ we obtain a reduced
two-component Landau theory
\begin{equation}
E=E_0+\frac{1}{2}\sum_\alpha m_\alpha\phi_\alpha^2+\frac{1}{4}\sum_{\alpha\beta}\Lambda_{\alpha\beta}\phi_\alpha^2\phi_\beta^2+\mathcal{O}(\phi^6),
\label{twofield}
\end{equation}
where $\alpha,\beta=\uparrow,m$, and the interspecies density-density
interaction has been renormalized to $\Lambda_{\uparrow
  m}=\lambda_{\uparrow m}-\gamma^2/m_\downarrow$ whilst
$\Lambda_{\alpha\alpha}=\lambda_{\alpha\alpha}$ remains unchanged. The
behavior of this Landau theory is governed by the sign and magnitude
of $\Lambda_{\uparrow m}$ \cite{Chaikin:Book}.
\label{Sect:Landau}
\begin{figure}
\includegraphics[clip,width=3.2in]{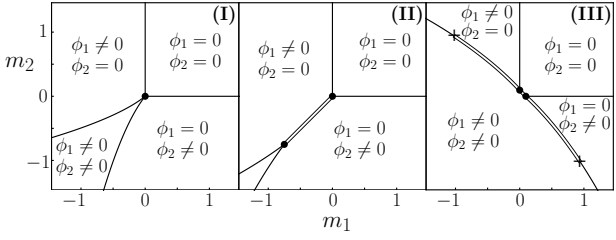}
\caption{Topologies of the reduced Landau theory (\ref{twofield}) with
  $\alpha,\beta\equiv 1,2$ and a term $(\phi_1^6+\phi_2^6)/6$ added
  for stability. We set $\Lambda_{11}=\Lambda_{22}=1$ and consider the
  evolution with $\Lambda_{12}$. We denote first order (continuous) transitions 
by double (single) lines. Junctions between phases are indicated by
  a dot, and the termination of first order lines by a cross. (I) $\Lambda_{12}^2<\Lambda_{11}\Lambda_{22}$: 
tetracritical point for $\Lambda_{12}=0.8$.
(II) $\Lambda_{12}>\sqrt{\Lambda_{11}\Lambda_{22}}$: 
first order transition between one-component condensates 
for $\Lambda_{12}=1.5$. (III) $\Lambda_{12}<-\sqrt{\Lambda_{11}\Lambda_{22}}$:
first order transition with condensation of both order parameters for 
$\Lambda_{12}=-1.5$.}
\label{Fig:Top}
\end{figure}
There are three distinct cases to consider as illustrated in
Fig.~\ref{Fig:Top}; (I) if $\Lambda_{\uparrow\uparrow}\Lambda_{m
  m}>\Lambda^2_{\uparrow m}$ this describes a tetracritical point;
(II) if $\Lambda_{\uparrow
  m}>\sqrt{\Lambda_{\uparrow\uparrow}\Lambda_{m m}}$ then there is a
first order transition between the two single component superfluids;
(III) if $\Lambda_{\uparrow
  m}<-\sqrt{\Lambda_{\uparrow\uparrow}\Lambda_{m m}}$ then the system
is unstable at fourth order indicating the presence of a first order
transition to condensation of both order parameters. For the
parameters used in Fig.~\ref{Fig:MFT} we find explicit examples of
types (I) and (III); see Figs.~\ref{Fig:Tip2} and \ref{Fig:Tip}.
At the lower tetracritical point in Fig.~\ref{Fig:MFT}(c), where
$m_\uparrow=m_m=0$, the Landau coefficients are given by
$\Lambda_{\uparrow \uparrow}=4\tilde{\epsilon}_\uparrow$,
\begin{equation}
  \Lambda_{mm } =\frac{4[g^2-\tilde{\epsilon}_m({\tilde\epsilon}_m-h)]}{h-\tilde{\epsilon}_m}\left[\frac{g^4+g^2(h-\tilde{\epsilon}_m)^2}{(h-2\tilde{\epsilon}_m)(h-\tilde{\epsilon}_m)^3}+1\right],
\end{equation}
and
\begin{equation}
  \Lambda_{\uparrow m} =
  -\frac{2g^2}{\tilde{\epsilon}_\downarrow}
  \left[
    \frac{g^2\tilde{\epsilon}_\downarrow
      +{\mathcal E}(\tilde{\epsilon}^2_\uparrow-\tilde{\epsilon}_\downarrow^2-2\tilde{\epsilon}_{\downarrow}\tilde{\epsilon}_{\uparrow})}
    {{\mathcal E}
      (\tilde{\epsilon}_\uparrow+\tilde{\epsilon}_{\downarrow})^2}
    +\frac{\tilde{\epsilon}_{\uparrow}}{\tilde{\epsilon}_{\downarrow}-\tilde{\epsilon}_{\uparrow}}\right],
\end{equation}
where $\mu_{\rm
  T}=(\epsilon_\downarrow+\epsilon_\uparrow)/2+(\nu-\sqrt{3g^2+\nu^2})/3$
for $2t_m=t_\downarrow=t_\uparrow=t$, $\nu \equiv
h+(\tilde{\epsilon}_\uparrow-\tilde{\epsilon}_\downarrow)/4$ and
$\mathcal{E}\equiv\tilde{\epsilon}_\uparrow+\tilde{\epsilon}_m$.
Setting $h=0$ as appropriate for Fig.~\ref{Fig:MFT}, the bifurcation
condition $\Lambda_{\uparrow
  m}=-\sqrt{\Lambda_{\uparrow\uparrow}\Lambda_{mm}}$ yields $\mu_{\rm
  D}\approx 0.203g+(\epsilon_\uparrow-\epsilon_\downarrow)/2$.  This
is in quantitative agreement with the bifurcation point shown in
Figs.~\ref{Fig:MFT} and \ref{Fig:Tip2}, where $g=1$. In a similar
fashion, for the parameters used in Fig.~\ref{Fig:MFT}, the $|-\rangle$ 
lobe tetracritical points bifurcate
at $\mu_{\rm D}\approx 0.445$ and $\mu_{\rm
  D}=(4/\sqrt{7}-1)^{1/2}\approx 0.715$; see
Figs.~\ref{Fig:Tip2} and \ref{Fig:Tip}.

\section{Magnetic Description}
\label{Sect:Magnetic}
\begin{figure}
  \includegraphics[width=2.5in]{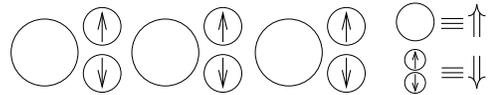}
  \caption{Schematic representation of the second Mott lobe with
    $\rho_{\rm T}=2$ showing the presence of a pair of atoms
    $\Downarrow$ or a molecule $\Uparrow$ at each site. XY exchange
    involves interchanging a molecule and a {\em pair} of atoms, 
 and is suppressed relative to the Ising interaction for small hoppings.}
\label{Fig:lobespins}
\end{figure}

Having discussed the phase diagram we turn our attention to the Mott
state with total density $\rho_{\rm T}=2$. This reveals Ising
transitions in both the heteronuclear and homonuclear lattice
problems \cite{Bhaseen:Feshising}.  To explore this second Mott lobe,
where the Feshbach term is operative, we adopt a magnetic description.
With a pair of atoms or a molecule at each site, we introduce
effective spins $|\!\Downarrow\rangle \equiv |1,1;0\rangle$,
$|\!\Uparrow\rangle \equiv |0,0;1\rangle$; see
Fig.~\ref{Fig:lobespins}.  The operators
\begin{equation}
S^+=a_m^\dagger a_\uparrow a_\downarrow, 
\quad 
S^-=a_\downarrow^\dagger a_\uparrow^\dagger a_m,
\quad
S^z=\frac{1}{2}\left(n_m-n_\downarrow n_\uparrow\right)
\label{hetspinrep}
\end{equation} 
or equally $S^z=[n_m-(n_\downarrow+n_\uparrow)/2]/2$, form a
representation of ${\rm su}(2)$ on this reduced Hilbert space.
Although the spin representation (\ref{hetspinrep}) contains three
bosons it is closely related to the more familiar Schwinger
construction \cite{Auerbach:Interacting}.  Deep within the Mott phase
we perform a strong coupling $t/U$ expansion \cite{Duan:Control}:
\begin{equation}
H= J\sum_{\langle ij\rangle}S_i^zS_j^z+\sum_i\left(hS_i^z+\Gamma
S_i^x\right),
\label{LTFI}
\end{equation}
where $\Gamma=2g$,
$h=\epsilon_m-\epsilon_\downarrow-\epsilon_\uparrow-V$, and the exchange 
interaction is given by
\begin{equation}
J=2\left(\frac{t_\downarrow^2+t_\uparrow^2}{U-V}+\frac{t_m^2}{2U}\right).
\label{JAFM}
\end{equation}
Here we focus on the antiferromagnetic case with $J>0$. In writing (\ref{LTFI}) we have omitted the constant,
$\tilde\epsilon_m-h/2-Jz/8$ per site which is necessary for
quantitative comparisons; see App. \ref{App:Ising}.  More generally,
in the presence of asymmetry between $U_{m\uparrow}$ and
$U_{m\downarrow}$
\begin{equation}
  J=2\left(\frac{t_\downarrow^2}{U_{m\downarrow}-U_{\uparrow\downarrow}}+\frac{t_\uparrow^2}{U_{m\uparrow}-U_{\uparrow\downarrow}}
    +\frac{t_m^2}{U_{m\downarrow}+U_{m\uparrow}}\right).
\end{equation}
The structure of this exchange is readily seen from the energy cost
for each individual hopping process in second order perturbation
theory.  The Hamiltonian~(\ref{LTFI}) takes the form of a quantum
Ising model in a longitudinal and transverse field.  The longitudinal
field, $h$, reflects the energetic asymmetry between a molecule
$|\!\!\Uparrow\rangle$, and a pair of atoms $|\!\!\Downarrow\rangle$.
The transverse field, $\Gamma\equiv 2g$, encodes Feshbach conversion and
induces quantum fluctuations in the ground state; 
see Fig.~\ref{Fig:lobespins}. In particular, in 
one dimension and with $h=0$, the model (\ref{LTFI}) is
exactly solvable by fermionization. It exhibits a quantum phase
transition at $\Gamma=J/2$ from an ordered to disordered phase
\cite{Lieb:Two,Schultz:2D,Pfeuty:TFI}. For $h\neq 0$, the location of
this transition is modified; see for example Ref.~\cite{Ovchinnikov:Mixed}.

\section{Numerical Simulations}
\label{Sect:Num}
The model~(\ref{LTFI}) plays an important role in quantum magnetism
and quantum phase transitions \cite{Sondhi:Continuous,Sachdev:QPT}.
To verify this realization in our {\em bosonic} model
(\ref{atmolham}), we perform exact diagonalization on the 1D quantum
system~(\ref{atmolham}) with periodic boundary conditions, at zero
temperature. The large Hilbert space $\propto 2^{3N}$ of the
multicomponent system under consideration restricts our simulations to
$N\leq8$ sites. In Sec.~\ref{Sect:Numpd} we present results for the
overall phase diagram in the grand canonical ensemble (allowing all
possible states) for $N=6$ sites, before moving on to a detailed
canonical ensemble study (allowing only states with fixed total
density $\rho_{\rm T}=2$) of the Ising quantum phase transition in
Sec.~\ref{Sect:Numising}. Comparing directly to results for the Ising
model at a given system size provides a good understanding of
finite-size effects in the bosonic problem.
\subsection{Grand Canonical Phase Diagram}
\label{Sect:Numpd}
\begin{figure}
  \includegraphics[width=3in]{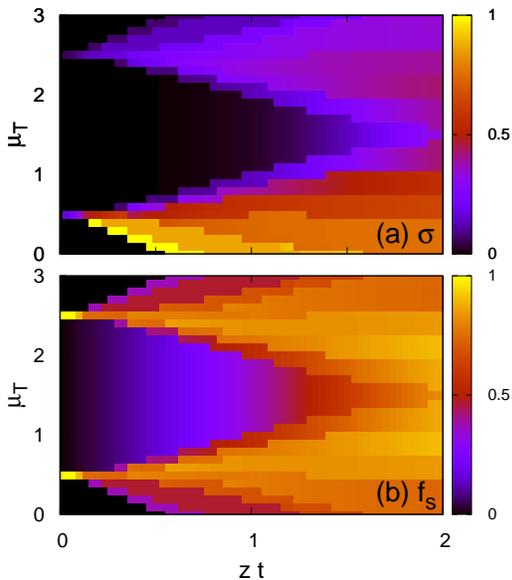}
  \caption{(Color online) (a) Density fluctuations and (b) superfluid
    fraction for the 1D bosonic model~(\ref{atmolham}) obtained by
    exact diagonalization for $N=6$ sites.  We set
    $\epsilon_\downarrow=\epsilon_\uparrow=1$, $\epsilon_m=2$, $U=1$,
    $V=0$, $t=t_\downarrow=t_\uparrow=2t_m$ and $\mu_{\rm D}=0.45$.
    The extent of the Mott lobes is in good agreement with
    the mean field predictions in Fig.~\ref{Fig:MFTUV}(a).  The system
    sizes are insufficient to resolve the quantum phase transitions
    between the distinct superfluids shown in Figs.~\ref{Fig:MFT} and
    \ref{Fig:MFTUV}.}
\label{Fig:Numpd}
\end{figure}
In Fig.~\ref{Fig:Numpd} 
we present results for the fluctuations in the total density, $\sigma
= \sqrt{\langle n_{\rm T}^2\rangle - \langle n_{\rm T}\rangle^2}$
where $n_\text{T}=\sum_\alpha n_{i\alpha}$ for an arbitrary site $i$,
and the total superfluid fraction
\begin{equation}
f_\text{s} = \frac{N}{\sum_\alpha t_\alpha \langle n_\alpha \rangle}
\frac{E_{\rm gs}(\Theta)-E_{\rm gs}(0)}{\Theta^2} \,.
\end{equation}
Here we impose a phase twist $\Theta\ll\pi$ and calculate the change
in the ground state energy $E_{\rm gs}(\Theta)$
\cite{Roth:Twocolor,Silver:BHC}. These key observables show the onset
of superfluidity, and the results are in good qualitative agreement
with each other, and with the mean field phase diagram shown in
Fig.~\ref{Fig:MFTUV}(a). As found in the single component Bose--Hubbard
model \cite{Fisher:Bosonloc}, the Mott lobes develop sharpened tips
due to enhanced fluctuations in one dimension
\cite{Kuhner:Phases,Kuhner:1DBH}.  However here, 
the Mott lobe 
remains symmetric around the mid-point due to the symmetry of 
the Hamiltonian (\ref{atmolham}) under particle-hole
and spin flip operations, $a_\alpha\leftrightarrow a_\alpha^\dagger$,
$\mu_{\rm T}\rightarrow \epsilon_m+U-\mu_{\rm T}$,
$a_\downarrow\leftrightarrow a_\uparrow$, for our chosen parameters with 
$t_\downarrow=t_\uparrow$ and $h=0$. The system sizes accessible
by exact diagonalization are insufficient to resolve the 
quantum phase transitions between the distinct superfluids 
shown in Figs.~\ref{Fig:MFT} and \ref{Fig:MFTUV}.  
However, as we will demonstrate 
in Sec.~\ref{Sect:Numising}, they provide
compelling evidence for the magnetic structure of the Mott phases
\cite{Bhaseen:Feshising}.

\subsection{Ising Transition in the Second Mott Lobe}
\label{Sect:Numising}
To employ greater system sizes we switch to the canonical ensemble
with fixed density, $\rho_{\rm T}=2$. To study the Ising transition we 
work in the region of small hopping parameters,
\begin{figure}
\includegraphics[width=3in]{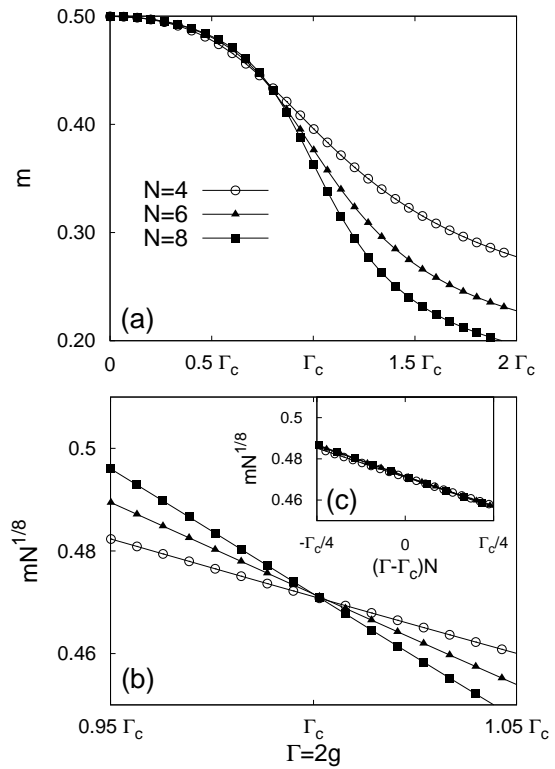}
\caption{(a) Pseudo staggered magnetization, $m$, versus $\Gamma=2g$,
  obtained from exact diagonalization of the 1D bosonic
  model~(\ref{atmolham}) for different system sizes, $N$, at density
  $\rho_{\rm T}=2$. We take $\epsilon_\downarrow=\epsilon_\uparrow=1$,
  $\epsilon_m=2$ (corresponding to $h=0$), $U=1$, $V=0$ and set
  $t=t_\downarrow=t_\uparrow=2t_m=0.01$ corresponding to
  $J=17t^2/4$. (b) Finite size rescaling of
  $mN^\beta$ versus $\Gamma=2g$ showing the presence of the Ising
  quantum phase transition at $\Gamma=\Gamma_\text{c}\approx
  2.08\times 10^{-4}$ and the critical exponent $\beta=1/8$ of the 2D
  classical Ising model.  The critical coupling $\Gamma_c\approx 0.49J$ 
is close to the small hopping Ising result, $\Gamma_c=J/2$. 
(c) Scaling collapse as a function of
  $(\Gamma-\Gamma_\text{c})N^{1/\nu}$ corresponding to $\nu=1$.}
\label{Fig:itrans}
\end{figure}
and begin with $h=0$ before exploring finite fields. Due to the
absence of spontaneous symmetry breaking in finite systems, the
staggered magnetization vanishes in the absence of an applied
staggered field. In view of this we focused our previous numerical
investigation \cite{Bhaseen:Feshising} on the pseudo staggered
magnetization, $m\equiv \langle |\sum_i (-1)^iS_i^z|\rangle/N$
\cite{Um:FiniteTFI}, where
$S_i^z=\left[n_{im}-(n_{i\uparrow}+n_{i\downarrow})/2\right]/2$ and
additional modulus signs are incorporated.  As shown in
Fig.~\ref{Fig:itrans}(a), this quantity is rendered finite.  Adopting
the finite-size scaling form, $m=N^{-\beta/\nu} \bar
m\left[(\Gamma-\Gamma_\text{c})N^{1/\nu}\right]$ \cite{Um:FiniteTFI},
we plot $mN^{1/8}$ versus $g$ for different system sizes, $N$, in
Fig.~\ref{Fig:itrans}(b). The curves cross close to the critical
coupling, $\Gamma_\text{c}=J/2$, of the purely transverse field Ising
model.  Moreover, the scaling collapse shown in
Fig.~\ref{Fig:itrans}(c) is consistent with the critical exponents
$\beta=1/8$ and $\nu=1$ for the 2D classical model. In spite of this
evident success, it is clearly desirable to examine this transition in
direct physical observables, and we turn our attention to this below.
In particular, we establish a direct connection with analytical
results and confirm additional Ising critical exponents.

\begin{figure}
\includegraphics[width=3in]{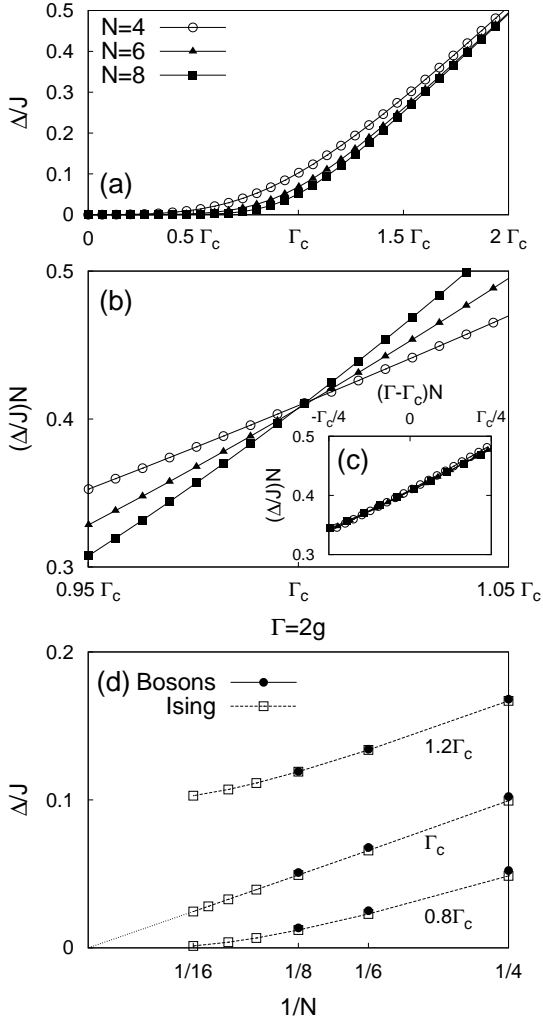}
\caption{(a) Dimensionless energy gap $\Delta/J$ versus
  $\Gamma=2g$ obtained by exact diagonalization of the 1D bosonic
  Hamiltonian~(\ref{atmolham}) for different system sizes, $N$.  We
  set $\epsilon_\downarrow=\epsilon_\uparrow=1$, $\epsilon_m=2$,
  $U=1$, $V=0$, $t=t_\downarrow=t_\uparrow=2t_m=0.01$. 
(b) Scaled energy gap $(\Delta/J)N^\nu$ versus
  $\Gamma=2g$ showing the presence of the Ising quantum phase
  transition at the numerically extracted value,
  $\Gamma_\text{c}\approx 2.08 \times 10^{-4}$, and the correlation
  length exponent $\nu=1$. (c)
  Scaling collapse as a function of
  $(\Gamma-\Gamma_\text{c})N^{1/\nu}$ corresponding to $\nu=1$. (d)
  The numerically extracted critical coupling, $\Gamma_\text{c}\approx
  2.08\times 10^{-4}$, obtained from the crossing point in
  Fig.~\ref{Fig:Gap}\,(b) is consistent with the gap closing in the
  thermodynamic limit, $N\rightarrow\infty$.  We set $\Gamma$ equal to
  multiples of $J/2$ (or the numerically extracted bosonic coupling,
  $\Gamma_\text{c}$) for the Ising (bosonic) model.}
\label{Fig:Gap}
\end{figure}

As shown in Fig.~\ref{Fig:Gap}, the correlation length exponent,
$\nu=1$, also follows directly from the gap data. Here we plot
$(\Delta/J)N^\nu$ versus $\Gamma=2g$, where $\Delta\equiv E_{\rm
  ex}-E_{\rm gs}$ is the energy gap between the first excited state
and the ground state.  As indicated in Fig.~\ref{Fig:Gap}(d),
\begin{figure}
\includegraphics[width=3in]{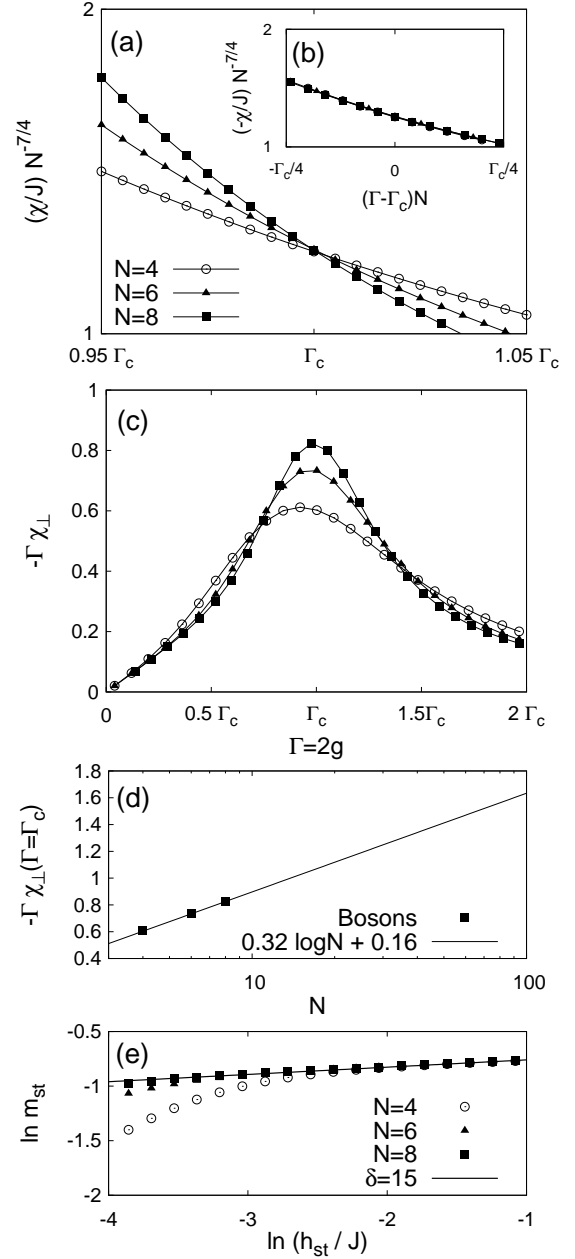}
\caption{(a) Staggered magnetic susceptibility obtained by exact
  diagonalization of the 1D bosonic Hamiltonian~(\ref{atmolham})
  revealing the Ising quantum phase transition at $\Gamma_\text{c}\approx
  2.08\times 10^{-4}$, and the critical exponent $\gamma=7/4$.  
We set $\epsilon_\downarrow=\epsilon_\uparrow=1$,
  $\epsilon_m=2$, $U=1$, $V=0$, $t=t_\downarrow=t_\uparrow=2t_m=0.01$.
(b) Scaling collapse showing $\nu=1$. (c) The transverse magnetic 
susceptibility, $\chi_\perp$, plays an analogous 
role to the specific heat capacity of the 2D classical 
Ising model. (d) The logarithmic divergence at the critical point 
yields $\alpha=0$ \cite{Um:FiniteTFI}. (e) 
$\log(m_\text{st})$ versus $\log(h_\text{st}/J)$ to extract
  the exponent $\delta=15$ for the bosonic model~(\ref{atmolham})
  using $m_\text{st}\sim h_{\rm st}^{1/\delta}$ for
  $\Gamma=\Gamma_\text{c}\approx 2.08\times 10^{-4}$.}
\label{Fig:chi}
\end{figure}
the values of the gap are also in very good quantitative agreement
with finite-size simulations of the Ising model. 

In a similar fashion
we obtain the exponent, $\gamma=7/4$, from the magnetic susceptibility
$\chi=\left.\partial m_{\rm st}/\partial h_{\rm
    st}\right|_{h\rightarrow 0^+}$, where $m_{\rm st}\equiv \langle
\sum_i (-1)^iS_i^z\rangle/N$ and $h_{\rm st}$ is a staggered magnetic
field applied to the bosonic Hamiltonian~(\ref{atmolham}), $\Delta
H=-h_{\rm st}\sum_i(-1)^iS_i^z$; see Figs.~\ref{Fig:chi}(a) and (b).
A significant advantage of this somewhat more involved procedure is
that the staggered magnetic field couples directly to the genuine
order parameter, $m_\text{st}$, without the need for modification.
Likewise, we compute the transverse susceptibility,
$\chi_\perp=\partial m_\perp/\partial\Gamma$, where $m_\perp=\sum_i
\langle S_i^x\rangle/N$ is the transverse magnetization.  Since the
transverse field acts to disorder the system, this plays a similar
role to the specific heat capacity of the 2D classical Ising model
\cite{Um:FiniteTFI}.  The dependence shown in Fig.~\ref{Fig:chi}(d) is
consistent with the Ising critical exponent $\alpha=0$. At
criticality, the relation, $m_\text{st}\sim h_\text{st}^{1/\delta}$,
is also furnished with $\delta=15$ as shown in
Fig.~\ref{Fig:chi}(e).

\begin{figure}
\includegraphics[width=3in]{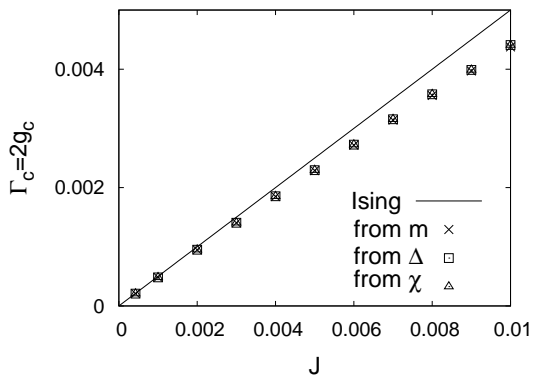}
\caption{Ising quantum phase transition in the bosonic
  model~(\ref{atmolham}) showing the emergence of the critical line,
  $\Gamma_c=J/2$, in the regime of small hoppings.  We set
  $\epsilon_\downarrow=\epsilon_\uparrow=1$, $\epsilon_m=2$, $U=1$,
  $V=0$, $t=t_\downarrow=t_\uparrow=2t_m$ and vary the hopping, $t$,
  corresponding to different values of $J=17t^2/4$. }
\label{Fig:lintrans}
\end{figure}

All of the above diagnostics confirm the consistency of the measured
Ising critical point \cite{Bhaseen:Feshising}, and we may track this
transition within the Mott lobe. The results in
Fig.~\ref{Fig:lintrans} show clear Ising behavior at small hoppings
corresponding to $\Gamma_\text{c}\approx J/2$.
With increasing hopping the boundary peels away from this linear slope
and we find $\Gamma_\text{c}<J/2$.  Additionally, enhanced finite-size
effects result in slightly different estimates of $\Gamma_\text{c}$
from $m$, $\Delta$ and $\chi$; see Fig.~\ref{Fig:lintrans}.  This will
be explored in future density matrix renormalization group (DMRG)
work \cite{Ejima:inprep}.

\begin{figure}
\includegraphics[width=3in]{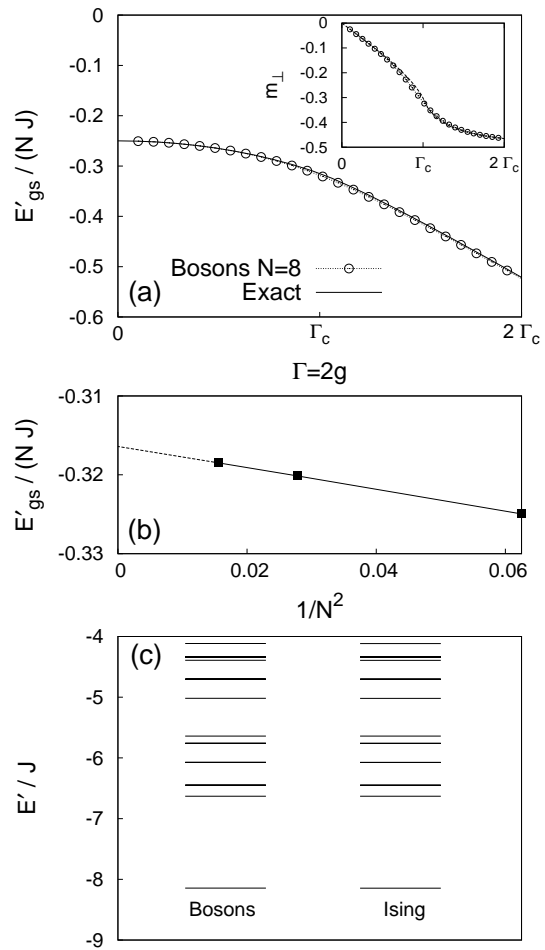}
\caption{Comparison of (a) the shifted ground state energy $E_{\rm
    gs}^\prime\equiv E_{\rm gs}-({\tilde\epsilon}_m-h/2-Jz/8)N$, and
  (Inset) transverse magnetization $m_\perp=\sum_i \langle
  S_i^x\rangle$/N of the bosonic model~(\ref{atmolham}) and the exact
  thermodynamic results (\ref{einf}) and~(\ref{sx}) for the Ising
  model. We set $\epsilon_\downarrow=\epsilon_\uparrow=1$,
  $\epsilon_m=2$, $U=1$, $V=0$,
  $t=t_\downarrow=t_\uparrow=2t_m=0.01$. (b) Finite size corrections
  for $E_{\rm gs}^\prime$ of the
  bosonic model~(\ref{atmolham}), with $\Gamma=\Gamma_\text{c}\approx
  2.08\times 10^{-4}$.  We use the finite size scaling result, $E_{\rm
    gs}^\prime/JN=e_\infty^\prime/J-\pi c (v/J)/6N^2$.  The intercept
  at -0.3164 is in good agreement with the exact Ising
  result~(\ref{einf}) in the thermodynamic limit,
  $e_\infty^\prime/J=-1/\pi\approx -0.3183$ when $\Gamma=J/2$.  The
  slope, $-\pi c/12$, yields the central charge, $c\approx 0.52$,
  where we use the characteristic velocity $v=J/2$. (c) 
Comparison of the low-energy spectra obtained by exact
  diagonalization for $N=8$ sites in the bosonic 
model~(\ref{atmolham}) with the same parameters as used in panel (a) with 
$\Gamma=2g=2J$, and the Ising model~(\ref{LTFI}) with $J=17t^2/4$, $h=0$, and
  $\Gamma=2J$.}
\label{Fig:Compare}
\end{figure}

As shown in Fig.~\ref{Fig:Compare}\,(a), the results obtained for
$N=8$ sites are in good agreement with exact results pertaining to the
thermodynamic limit of the quantum Ising model for both the
ground-state energy density \cite{Pfeuty:TFI}
\begin{equation}
e_\infty =-\frac{1}{4\pi}\int_0^\pi dk \sqrt{4\Gamma^2+J^2+4\Gamma J\cos k},
\label{einf}
\end{equation}
and the transverse magnetization \cite{Pfeuty:TFI,Hieida:Reentrant} 
\begin{equation}
\langle S_i^x\rangle=-\int_0^\pi \frac{dk}{2\pi}
\frac{2\Gamma+J\cos k}{\sqrt{4\Gamma^2+J^2+4\Gamma J\cos k}}.
\label{sx}
\end{equation}
In this comparison we define $E_{\rm gs}^\prime\equiv E_{\rm
  gs}-({\tilde\epsilon}_m-h/2-Jz/8)N$ for the bosonic system
(\ref{atmolham}) in order to take into account the constant offset in
the mapping (\ref{LTFI}).
The residual finite size effects may be used to find the central
charge of the bosonic system.  With periodic boundary conditions the
ground state energy depends on the system size, $L=Na$, according to
$E_{\rm gs}=e_\infty L-\pi cv/6L+\dots$, where $e_\infty$ is the
ground state energy density in the thermodynamic limit, $v$ is the
effective velocity in the linearized low energy dispersion, and $c$ is
the central charge
\cite{Blote:Conformal,Affleck:Universal,Cardy:scal}. For the 1D Ising
model~(\ref{LTFI}), the dispersion relation may be obtained by
fermionization \cite{Lieb:Two,Schultz:2D,Pfeuty:TFI}. This yields the
characteristic velocity, $v=\left.\partial\varepsilon_k/\partial
  k\right|_{k=0}=J/2$.  The recovery of the Ising model central
charge, $c=1/2$ \cite{BPZ,Francesco:CFT}, is shown in
Fig.~\ref{Fig:Compare}(b).  Although the majority of our discussion
has been on the ground state properties, and the first excited state,
the agreement between the Ising model description and the bosonic
Hamiltonian also extends to the low-energy spectrum shown in
Fig.~\ref{Fig:Compare}(c).

\begin{figure}
\includegraphics[clip,width=3in]{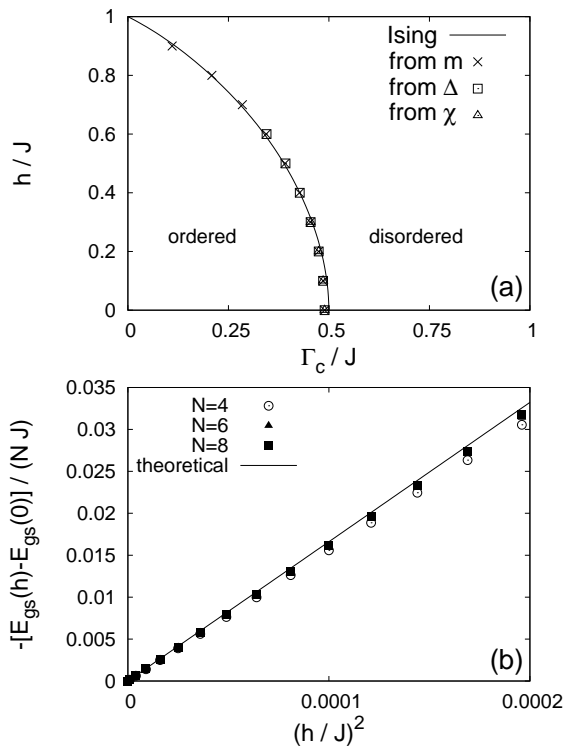}
\caption{(a) Evolution of the Ising quantum phase transition 
with the longitudinal field, $h$, in the 1D bosonic model 
(\ref{atmolham}). Here $\epsilon_m=2+h$ and all other 
parameters as in Fig.~\ref{Fig:itrans}. (b) 
Quadratic correction to the ground state energy for the
    bosonic Hamiltonian~(\ref{atmolham}) due to a longitudinal
    magnetic field, $h$, in the vicinity of the $h=0$ Ising critical
    point. We set $\epsilon_\downarrow=\epsilon_\uparrow=1$,
    $\epsilon_m=2$, $U=1$, $V=0$,
    $t=t_\downarrow=t_\uparrow=2t_m=0.01$ and
    $\Gamma=\Gamma_\text{c}\approx 2.08\times 10^{-4}$.  The results
    are close to the theoretical value, $\delta E_{\rm
      gs}^{(2)}\approx -0.0706h^2 N$, at $\Gamma=\Gamma_c$
    \cite{Ovchinnikov:Mixed}.}
\label{Fig:Isingh}
\end{figure}

Having provided a thorough description of the Ising quantum phase
transition in the absence of a magnetic field, let us now consider its
effect.  In the presence of a finite longitudinal field, $h$, the
location of the Ising critical point is modified as shown in
Fig.~\ref{Fig:Isingh}(a).
The results obtained from the pseudo staggered
magnetization, magnetic susceptibility, and gap data are in good
agreement with exact diagonalization and DMRG results \cite{Ovchinnikov:Mixed} 
for the Ising model. Finite-size effects
increase with increasing magnetic field, and for $h/J\gtrsim0.6$ and
$N=4,6,8$, clean scaling is observed for $m$, but not for $\Delta$
and $\chi$. In addition, departures from the $h=0$ critical point 
yield the expected quadratic dependence of the ground state energy 
on the longitudinal field, $h$. The relation 
$\delta E_{\rm gs}^{(2)}=-0.0706h^2N$ 
\cite{Ovchinnikov:Mixed,Muller:Sus} 
is recovered in Fig.~\ref{Fig:Isingh}(b). 
These results show the persistence of an Ising quantum phase transition
in the bosonic model without the need to fine tune to zero magnetic field.

\section{Conclusions}
\label{Sect:Conc}
In this work we have provided a detailed study of bosonic
heteronuclear mixtures with Feshbach resonant pairing in optical
lattices. The model displays a rich phase diagram with an intricate
network of quantum critical points and phase transitions, and we have
anchored this behavior to the predictions of a reduced two-component
Landau theory. We have substantiated our previous findings of an Ising
quantum phase transition occurring within the second Mott lobe
\cite{Bhaseen:Feshising} by significantly extending the range of
physical observables.  In particular, we have confirmed the additional
Ising critical exponents, $\alpha=0$, $\gamma=7/4$, and $\delta=15$,
for the one-dimensional bosonic system.  There are many directions for
further investigation including the superfluid properties, and studies
away from the hardcore limit.  Cold atoms in optical lattices may
represent ideal systems in which to explore magnetization
distributions \cite{Lamacraft:OPS} and quantum quenches
\cite{Rossini:Thermal,Calabrese:Quench} in quantum Ising models.

\begin{acknowledgments}
  We are grateful to F. Essler, S. Ejima and H. Fehske for helpful
  discussions. MJB, AOS, and BDS acknowledge EPSRC grant
  no. EP/E018130/1. MH acknowledges the hospitality of the TCM group
  at the University of Cambridge.
\end{acknowledgments}

\appendix

\section{Landau Coefficients}
\label{App:Landau}
To calculate the phase boundaries delimiting the single component
superfluids from the Mott lobes it is sufficient to examine where the
relevant mass terms change sign.  In particular, for the second Mott
lobe, $|-\rangle$, we find
\begin{equation}
E_-=E_-^0+\frac{1}{2}\sum_\nu m_\nu\phi_\nu^2+\dots
\end{equation}
where
\begin{equation}
m_m=zt_m\left\{1-zt_m\frac{\chi(\chi+U)+\frac{h}{2}(\frac{h}{2}-{\tilde\epsilon}_m-U)}
{\chi\left[(\chi+U)^2-(\frac{h}{2}-{\tilde\epsilon}_m-U)^2\right]}\right\},
\end{equation}
and
\begin{equation}
  m_{\uparrow\downarrow}=zt_{\uparrow\downarrow}\left\{1-zt_{\uparrow\downarrow}
    \frac{\chi(\chi-\beta)+\frac{h}{2}({\tilde\epsilon}_{\uparrow\downarrow}+\alpha)}
    {\chi\left[(\chi-\beta)^2-({\tilde\epsilon}_{\uparrow\downarrow}+\alpha)^2\right]}\right\}.
\end{equation}
Here we denote $\chi=\sqrt{g^2+(h/2)^2}$, 
$\alpha=h/2+(U+V)/2$ and $\beta=(V-U)/2$. The coefficients for 
$|0,0;0\rangle$ read
\begin{align}
m_{\uparrow\downarrow}(0,0;0)&=zt_{\uparrow\downarrow}\left(1-\frac{zt_{\uparrow\downarrow}}{\tilde{\epsilon}_{\uparrow\downarrow}}\right),
\label{mdu000}\\
m_m(0,0;0)&=zt_m\left(1-\frac{zt_m(h-\tilde{\epsilon}_m)}{\chi^2-(\tilde{\epsilon}_m-\frac{h}{2})^2}\right),
\label{mm000}
\end{align}
whilst those pertaining to $|1,0;0\rangle$ read
\begin{align}
m_m(1,0;0)&=zt_m\left(1-\frac{zt_m}{\tilde{\epsilon}_m+U}\right),\label{mm100}
\\
m_{\downarrow}(1,0;0)&=zt_\downarrow\left(1+\frac{zt_\downarrow}{\tilde{\epsilon}_\downarrow}\right),\label{md100}
\\
m_{\uparrow}(1,0;0)&=zt_\uparrow\left(1+\frac{h+\tilde{\epsilon}_\uparrow+V}{\chi^2-(\frac{h}{2}+\tilde{\epsilon}_\uparrow+V)^2}\right).
\label{mu100}
\end{align}
The remaining coefficients may be obtained by interchanging particles 
and holes and ups and downs as appropriate. For example, 
the coefficients $m_\alpha(0,1;0)$ are obtained by interchanging $\downarrow\leftrightarrow \uparrow$ in equations (\ref{mm100}) --- (\ref{mu100}).
Likewise, the coefficients $m_\alpha(1,1;1)$ are obtained by the particle--hole
transformation, $\tilde{\epsilon}_{\uparrow\downarrow}\rightarrow
-(\tilde{\epsilon}_{\uparrow\downarrow}+U+V)$ and $\tilde{\epsilon}_m
\rightarrow -(\tilde{\epsilon}_m+2U)$, on equations (\ref{mdu000}) and (\ref{mm000}).

\section{Derivation of the Ising Hamiltonian}
\label{App:Ising}

In this Appendix we derive the effective spin model describing the
second Mott lobe of the heteronuclear Hamiltonian.  Employing the spin
operators as defined in equation~(\ref{hetspinrep}) and noting the
form of the zero hopping eigenstates we may re-write the bosonic
Hamiltonian~(\ref{atmolham}) as
\begin{equation}
H=NE_0\mathcal{P}_0+H^0+H^\prime,
\end{equation}
where $N$ is the number of lattice sites,
\begin{equation}
  H^0=\left(\sum_{i\alpha}\tilde{\epsilon}_\alpha n_{i\alpha}+\sum_{i\alpha\alpha^\prime}\frac{U_{\alpha\alpha^\prime}}{2}:n_{i\alpha}n_{i\alpha^\prime}:\right)(1-\mathcal{P}_0),
\end{equation}
and
\begin{equation}
  H^\prime=\sum_{i,\sigma=\pm}\sigma\chi|\sigma_i\rangle\langle\sigma_i|-\sum_{\langle ij \rangle,\alpha}t_\alpha(a^\dagger_{i\alpha}a_{j\alpha}+\text{h.c.}).
\end{equation}
Here, $E_0=\tilde{\epsilon}_m-h/2$,
and $\mathcal{P}_0=\prod_i |\!\!\Downarrow_i\rangle\langle\Downarrow_i\!\!|
+|\!\!\Uparrow_i\rangle\langle\Uparrow_i\!\!|$
is the projection operator
onto the quasi-degenerate subspace of a molecule or two atoms on each
site. The effective Hamiltonian within this subspace
is \cite{Messiah:QM}
\begin{equation}
  H_S=NE_0\mathcal{P}_0+\mathcal{P}_0H^\prime\mathcal{P}_0
  +\mathcal{P}_0H^\prime {S^\prime}^1 H^\prime\mathcal{P}_0
  +\mathcal{O}({H^\prime}^3),
\end{equation}
where ${S^\prime}^1=\sum_{m\neq \Uparrow,\Downarrow}|m\rangle\langle
m|/(NE_0-E_{|m\rangle})$. Since only the hopping contribution to $H^\prime$
connects this subspace to the other states one immediately finds
\begin{equation}
\begin{split}
  \mathcal{P}_0H^\prime\mathcal{P}_0&=\mathcal{P}_0\sum_{i,\sigma=\pm}\sigma\chi|\sigma_i\rangle\langle\sigma_i|\mathcal{P}_0\\
  &=\mathcal{P}_0\sum_i\left[g(S^+_i+S^-_i)+hS^z_i\right]\mathcal{P}_0.
\end{split}
\end{equation}
The only process contributing at second order is a particle hopping to
a neighboring site and back:
\begin{equation}
\begin{split}
  \mathcal{P}_0H^\prime {S^\prime}^1 H^\prime\mathcal{P}_0&=\sum_{\alpha,\langle ij \rangle}t^2_\alpha\mathcal{P}_0a^\dagger_{i\alpha}a_{j\alpha}{S^\prime}^1a_{i\alpha}a^\dagger_{j\alpha}\mathcal{P}_0\\
  &\qquad+(i\leftrightarrow j).
\end{split}
\end{equation}
Noting that
\begin{align}
  a^\dagger_{i m}a_{j m}\mathcal{P}_0&= |1,1;1\rangle_i|0,0;0\rangle_j\langle \Uparrow_j \Downarrow_i\!|\mathcal{P}_0,\\
  a^\dagger_{i \uparrow}a_{j \uparrow}\mathcal{P}_0&= |0,1;1\rangle_i|1,0;0\rangle_j\langle \Downarrow_j \Uparrow_i\!|\mathcal{P}_0,\\
  a^\dagger_{i \downarrow}a_{j \downarrow}\mathcal{P}_0& =
  |1,0;1\rangle_i|0,1;0\rangle_j\langle \Downarrow_j
  \Uparrow_i\!|\mathcal{P}_0,
\end{align}
one obtains
\begin{equation}
\begin{split}
  & \mathcal{P}_0H^\prime {S^\prime}^1 H^\prime\mathcal{P}_0=\\
  &\mathcal{P}_0\sum_{\langle ij \rangle}
  \left(-\frac{t_m^2}{2U}+\frac{t_\downarrow^2+t^2_\downarrow}{V-U}\right)|\Uparrow_j
  \Downarrow_i\rangle\langle \Uparrow_j \Downarrow_i|\mathcal{P}_0
  +(i\leftrightarrow j),
\end{split}
\end{equation}
where $U_{m\uparrow}=U_{m\downarrow}=U$ and
$U_{\uparrow\downarrow}=V$.  Using the identity
\begin{equation}
|\!\Uparrow_j \Downarrow_i\rangle\langle \Uparrow_j \Downarrow_i\!|=(1/2+S^z_j)(1/2-S^z_i)
\end{equation}
one obtains
\begin{equation}
\begin{split}
\mathcal{P}_0H^\prime {S^\prime}^1 H^\prime\mathcal{P}_0&=-\mathcal{P}_0J\sum_{\langle ij \rangle} \left(\frac{1}{4}-S^z_iS^z_j\right)\mathcal{P}_0,
\end{split}
\end{equation}
where $J$ is given by equation~(\ref{JAFM}). We conclude that 
\begin{equation}
\begin{split}
H&=\mathcal{P}_0\left[\sum_i\left(\tilde{\epsilon}_m-\frac{h}{2}-\frac{zJ}{8}\right)+J\sum_{\langle ij \rangle} S^z_iS^z_j\right.\\
&\left. \hspace{2cm} +\Gamma\sum_{i}S^x_i+h\sum_{i}S_i^z\right]\mathcal{P}_0,
\label{appisingham}
\end{split}
\end{equation}
as stated in equation~(\ref{LTFI}). In the zero hopping limit the
eigenstates of Hamiltonians~(\ref{atmolham}) and~(\ref{appisingham})
coincide, including the mixing induced by the Feshbach coupling.


\begin{thebibliography}{10}%
\makeatletter
\providecommand \@ifxundefined [1]{%
 \ifx #1\undefined \expandafter \@firstoftwo
 \else \expandafter \@secondoftwo
\fi
}%
\providecommand \@ifnum [1]{%
 \ifnum #1\expandafter \@firstoftwo
 \else \expandafter \@secondoftwo
\fi
}%
\providecommand \enquote [1]{``#1''}%
\providecommand \bibnamefont  [1]{#1}%
\providecommand \bibfnamefont [1]{#1}%
\providecommand \citenamefont [1]{#1}%
\providecommand\href[0]{\@sanitize\@href}%
\providecommand\@href[1]{\endgroup\@@startlink{#1}\endgroup\@@href}%
\providecommand\@@href[1]{#1\@@endlink}%
\providecommand \@sanitize [0]{\begingroup\catcode`\&12\catcode`\#12\relax}%
\@ifxundefined \pdfoutput {\@firstoftwo}{%
 \@ifnum{\z@=\pdfoutput}{\@firstoftwo}{\@secondoftwo}%
}{%
 \providecommand\@@startlink[1]{\leavevmode}%
 \providecommand\@@endlink[0]{}%
}{%
 \providecommand\@@startlink[1]{%
  \leavevmode
  \pdfstartlink
   attr{/Border[0 0 1 ]/H/I/C[0 1 1]}%
   user{/Subtype/Link/A<</Type/Action/S/URI/URI(#1)>>}%
  \relax
 }%
 \providecommand\@@endlink[0]{\pdfendlink}%
}%
\providecommand \url  [0]{\begingroup\@sanitize \@url }%
\providecommand \@url [1]{\endgroup\@href {#1}{\urlprefix}}%
\providecommand \urlprefix [0]{URL }%
\providecommand \Eprint[0]{\href }%
\@ifxundefined \urlstyle {%
  \providecommand \doi [1]{doi:\discretionary{}{}{}#1}%
}{%
  \providecommand \doi [0]{doi:\discretionary{}{}{}\begingroup
  \urlstyle{rm}\Url }%
}%
\providecommand \doibase [0]{http://dx.doi.org/}%
\providecommand \Doi[1]{\href{\doibase#1}}%
\providecommand \bibAnnote [3]{%
  \BibitemShut{#1}%
  \begin{quotation}\noindent
    \textsc{Key:}\ #2\\\textsc{Annotation:}\ #3%
  \end{quotation}%
}%
\providecommand \bibAnnoteFile [2]{%
  \IfFileExists{#2}{\bibAnnote {#1} {#2} {\input{#2}}}{}%
}%
\providecommand \typeout [0]{\immediate \write \m@ne }%
\providecommand \selectlanguage [0]{\@gobble}%
\providecommand \bibinfo [0]{\@secondoftwo}%
\providecommand \bibfield [0]{\@secondoftwo}%
\providecommand \translation [1]{[#1]}%
\providecommand \BibitemOpen[0]{}%
\providecommand \bibitemStop [0]{}%
\providecommand \bibitemNoStop [0]{.\EOS\space}%
\providecommand \EOS [0]{\spacefactor3000\relax}%
\providecommand \BibitemShut [1]{\csname bibitem#1\endcsname}%
%</preamble>
\bibitem{Santos:Dipolar}%
  \BibitemOpen
  \bibfield{author}{%
  \bibinfo {author} {\bibfnamefont{L.}~\bibnamefont{Santos}}, \bibinfo {author}
  {\bibfnamefont{G.~V.}\ \bibnamefont{Shlyapnikov}}, \bibinfo {author}
  {\bibfnamefont{P.}~\bibnamefont{Zoller}},\ and\ \bibinfo {author}
  {\bibfnamefont{M.}~\bibnamefont{Lewenstein}},\ }%
  \bibfield{journal}{%
  \bibinfo {journal} {Phys. Rev. Lett.}\ }%
  \textbf{\bibinfo {volume} {85}},\ \bibinfo {pages} {1791} (\bibinfo {year}
  {2000})%
  \bibAnnoteFile{NoStop}{Santos:Dipolar}%
\bibitem{Baranov:Dipolar}%
  \BibitemOpen
  \bibfield{author}{%
  \bibinfo {author} {\bibfnamefont{M.~A.}\ \bibnamefont{Baranov}},\ }%
  \bibfield{journal}{%
  \bibinfo {journal} {Phys. Rep.}\ }%
  \textbf{\bibinfo {volume} {464}},\ \bibinfo {pages} {71} (\bibinfo {year}
  {2008})%
  \bibAnnoteFile{NoStop}{Baranov:Dipolar}%
\bibitem{Ospelkaus:Reactions}%
  \BibitemOpen
  \bibfield{author}{%
  \bibinfo {author} {\bibfnamefont{S.}~\bibnamefont{Ospelkaus}}, \bibinfo
  {author} {\bibfnamefont{K.-K.}\ \bibnamefont{Ni}}, \bibinfo {author}
  {\bibfnamefont{D.}~\bibnamefont{Wang}}, \bibinfo {author}
  {\bibfnamefont{M.~H.~G.}\ \bibnamefont{de~Miranda}}, \bibinfo {author}
  {\bibfnamefont{B.}~\bibnamefont{Neyenhuis}}, \bibinfo {author}
  {\bibfnamefont{G.}~\bibnamefont{Qu{\'e}m{\'e}ner}}, \bibinfo {author}
  {\bibfnamefont{P.~S.}\ \bibnamefont{Julienne}}, \bibinfo {author}
  {\bibfnamefont{J.~L.}\ \bibnamefont{Bohn}}, \bibinfo {author}
  {\bibfnamefont{D.~S.}\ \bibnamefont{Jin}},\ and\ \bibinfo {author}
  {\bibfnamefont{J.}~\bibnamefont{Ye}},\ }%
  \bibfield{journal}{%
  \bibinfo {journal} {Science}\ }%
  \textbf{\bibinfo {volume} {327}},\ \bibinfo {pages} {853} (\bibinfo {year}
  {2010})%
  \bibAnnoteFile{NoStop}{Ospelkaus:Reactions}%
\bibitem{Papp:Hetero}%
  \BibitemOpen
  \bibfield{author}{%
  \bibinfo {author} {\bibfnamefont{S.~B.}\ \bibnamefont{Papp}}\ and\ \bibinfo
  {author} {\bibfnamefont{C.~E.}\ \bibnamefont{Wieman}},\ }%
  \bibfield{journal}{%
  \bibinfo {journal} {Phys. Rev. Lett.}\ }%
  \textbf{\bibinfo {volume} {97}},\ \bibinfo {pages} {180404} (\bibinfo {year}
  {2006})%
  \bibAnnoteFile{NoStop}{Papp:Hetero}%
\bibitem{Papp:Tunable}%
  \BibitemOpen
  \bibfield{author}{%
  \bibinfo {author} {\bibfnamefont{S.~B.}\ \bibnamefont{Papp}}, \bibinfo
  {author} {\bibfnamefont{J.~M.}\ \bibnamefont{Pino}},\ and\ \bibinfo {author}
  {\bibfnamefont{C.~E.}\ \bibnamefont{Wieman}},\ }%
  \bibfield{journal}{%
  \bibinfo {journal} {Phys. Rev. Lett.}\ }%
  \textbf{\bibinfo {volume} {101}},\ \bibinfo {pages} {040402} (\bibinfo {year}
  {2008})%
  \bibAnnoteFile{NoStop}{Papp:Tunable}%
\bibitem{Ticknor:Multiplet}%
  \BibitemOpen
  \bibfield{author}{%
  \bibinfo {author} {\bibfnamefont{C.}~\bibnamefont{Ticknor}}, \bibinfo
  {author} {\bibfnamefont{C.~A.}\ \bibnamefont{Regal}}, \bibinfo {author}
  {\bibfnamefont{D.~S.}\ \bibnamefont{Jin}},\ and\ \bibinfo {author}
  {\bibfnamefont{J.~L.}\ \bibnamefont{Bohn}},\ }%
  \bibfield{journal}{%
  \bibinfo {journal} {Phys. Rev. A}\ }%
  \textbf{\bibinfo {volume} {69}},\ \bibinfo {pages} {042712} (\bibinfo {year}
  {2004})%
  \bibAnnoteFile{NoStop}{Ticknor:Multiplet}%
\bibitem{Weber:Assoc}%
  \BibitemOpen
  \bibfield{author}{%
  \bibinfo {author} {\bibfnamefont{C.}~\bibnamefont{Weber}}, \bibinfo {author}
  {\bibfnamefont{G.}~\bibnamefont{Barontini}}, \bibinfo {author}
  {\bibfnamefont{J.}~\bibnamefont{Catani}}, \bibinfo {author}
  {\bibfnamefont{G.}~\bibnamefont{Thalhammer}}, \bibinfo {author}
  {\bibfnamefont{M.}~\bibnamefont{Inguscio}},\ and\ \bibinfo {author}
  {\bibfnamefont{F.}~\bibnamefont{Minardi}},\ }%
  \bibfield{journal}{%
  \bibinfo {journal} {Phys. Rev. A}\ }%
  \textbf{\bibinfo {volume} {78}},\ \bibinfo {pages} {061601(R)} (\bibinfo
  {year} {2008})%
  \bibAnnoteFile{NoStop}{Weber:Assoc}%
\bibitem{Thalhammer:Double}%
  \BibitemOpen
  \bibfield{author}{%
  \bibinfo {author} {\bibfnamefont{G.}~\bibnamefont{Thalhammer}}, \bibinfo
  {author} {\bibfnamefont{G.}~\bibnamefont{Barontini}}, \bibinfo {author}
  {\bibfnamefont{L.~D.}\ \bibnamefont{Sarlo}}, \bibinfo {author}
  {\bibfnamefont{J.}~\bibnamefont{Catani}}, \bibinfo {author}
  {\bibfnamefont{F.}~\bibnamefont{Minardi}},\ and\ \bibinfo {author}
  {\bibfnamefont{M.}~\bibnamefont{Inguscio}},\ }%
  \bibfield{journal}{%
  \bibinfo {journal} {Phys. Rev. Lett.}\ }%
  \textbf{\bibinfo {volume} {100}},\ \bibinfo {pages} {210402} (\bibinfo {year}
  {2008})%
  \bibAnnoteFile{NoStop}{Thalhammer:Double}%
\bibitem{Thalhammer:Coll}%
  \BibitemOpen
  \bibfield{author}{%
  \bibinfo {author} {\bibfnamefont{G.}~\bibnamefont{Thalhammer}}, \bibinfo
  {author} {\bibfnamefont{G.}~\bibnamefont{Barontini}}, \bibinfo {author}
  {\bibfnamefont{J.}~\bibnamefont{Catani}}, \bibinfo {author}
  {\bibfnamefont{F.}~\bibnamefont{Rabatti}}, \bibinfo {author}
  {\bibfnamefont{C.}~\bibnamefont{Weber}}, \bibinfo {author}
  {\bibfnamefont{A.}~\bibnamefont{Simoni}}, \bibinfo {author}
  {\bibfnamefont{F.}~\bibnamefont{Minardi}},\ and\ \bibinfo {author}
  {\bibfnamefont{M.}~\bibnamefont{Inguscio}},\ }%
  \bibfield{journal}{%
  \bibinfo {journal} {New J. Phys.}\ }%
  \textbf{\bibinfo {volume} {11}},\ \bibinfo {pages} {055044} (\bibinfo {year}
  {2009})%
  \bibAnnoteFile{NoStop}{Thalhammer:Coll}%
\bibitem{Barontini:Obs}%
  \BibitemOpen
  \bibfield{author}{%
  \bibinfo {author} {\bibfnamefont{G.}~\bibnamefont{Barontini}}, \bibinfo
  {author} {\bibfnamefont{C.}~\bibnamefont{Weber}}, \bibinfo {author}
  {\bibfnamefont{F.}~\bibnamefont{Rabatti}}, \bibinfo {author}
  {\bibfnamefont{G.}~\bibnamefont{Catani}}, \bibinfo {author}
  {\bibfnamefont{G.}~\bibnamefont{Thalhammer}}, \bibinfo {author}
  {\bibfnamefont{M.}~\bibnamefont{Inguscio}},\ and\ \bibinfo {author}
  {\bibfnamefont{F.}~\bibnamefont{Minardi}},\ }%
  \bibfield{journal}{%
  \bibinfo {journal} {Phys. Rev. Lett.}\ }%
  \textbf{\bibinfo {volume} {103}},\ \bibinfo {pages} {043201} (\bibinfo {year}
  {2009})%
  \bibAnnoteFile{NoStop}{Barontini:Obs}%
\bibitem{Catani:Entropy}%
  \BibitemOpen
  \bibfield{author}{%
  \bibinfo {author} {\bibfnamefont{J.}~\bibnamefont{Catani}}, \bibinfo {author}
  {\bibfnamefont{G.}~\bibnamefont{Barontini}}, \bibinfo {author}
  {\bibfnamefont{G.}~\bibnamefont{Lamporesi}}, \bibinfo {author}
  {\bibfnamefont{F.}~\bibnamefont{Rabatti}}, \bibinfo {author}
  {\bibfnamefont{G.}~\bibnamefont{Thalhammer}}, \bibinfo {author}
  {\bibfnamefont{F.}~\bibnamefont{Minardi}}, \bibinfo {author}
  {\bibfnamefont{S.}~\bibnamefont{Stringari}},\ and\ \bibinfo {author}
  {\bibfnamefont{M.}~\bibnamefont{Inguscio}},\ }%
  \bibfield{journal}{%
  \bibinfo {journal} {Phys. Rev. Lett.}\ }%
  \textbf{\bibinfo {volume} {103}},\ \bibinfo {pages} {140401} (\bibinfo {year}
  {2009})%
  \bibAnnoteFile{NoStop}{Catani:Entropy}%
\bibitem{Lamporesi:Mixed}%
  \BibitemOpen
  \bibfield{author}{%
  \bibinfo {author} {\bibfnamefont{G.}~\bibnamefont{Lamporesi}}, \bibinfo
  {author} {\bibfnamefont{J.}~\bibnamefont{Catani}}, \bibinfo {author}
  {\bibfnamefont{G.}~\bibnamefont{Barontini}}, \bibinfo {author}
  {\bibfnamefont{Y.}~\bibnamefont{Nishida}}, \bibinfo {author}
  {\bibfnamefont{M.}~\bibnamefont{Inguscio}},\ and\ \bibinfo {author}
  {\bibfnamefont{F.}~\bibnamefont{Minardi}},\ }%
  \bibfield{journal}{%
  \bibinfo {journal} {Phys. Rev. Lett.}\ }%
  \textbf{\bibinfo {volume} {104}},\ \bibinfo {pages} {153202} (\bibinfo {year}
  {2010})%
  \bibAnnoteFile{NoStop}{Lamporesi:Mixed}%
\bibitem{Simoni:Near}%
  \BibitemOpen
  \bibfield{author}{%
  \bibinfo {author} {\bibfnamefont{A.}~\bibnamefont{Simoni}}, \bibinfo {author}
  {\bibfnamefont{M.}~\bibnamefont{Zaccanti}}, \bibinfo {author}
  {\bibfnamefont{C.}~\bibnamefont{D'Errico}}, \bibinfo {author}
  {\bibfnamefont{M.}~\bibnamefont{Fattori}}, \bibinfo {author}
  {\bibfnamefont{G.}~\bibnamefont{Roati}}, \bibinfo {author}
  {\bibfnamefont{M.}~\bibnamefont{Inguscio}},\ and\ \bibinfo {author}
  {\bibfnamefont{G.}~\bibnamefont{Modugno}},\ }%
  \bibfield{journal}{%
  \bibinfo {journal} {Phys. Rev. A}\ }%
  \textbf{\bibinfo {volume} {77}},\ \bibinfo {pages} {052705} (\bibinfo {year}
  {2008})%
  \bibAnnoteFile{NoStop}{Simoni:Near}%
\bibitem{Pilch:Obs}%
  \BibitemOpen
  \bibfield{author}{%
  \bibinfo {author} {\bibfnamefont{K.}~\bibnamefont{Pilch}}, \bibinfo {author}
  {\bibfnamefont{A.~D.}\ \bibnamefont{Lange}}, \bibinfo {author}
  {\bibfnamefont{A.}~\bibnamefont{Prantner}}, \bibinfo {author}
  {\bibfnamefont{G.}~\bibnamefont{Kerner}}, \bibinfo {author}
  {\bibfnamefont{F.}~\bibnamefont{Ferlaino}}, \bibinfo {author}
  {\bibfnamefont{H.-C.}\ \bibnamefont{N\"agerl}},\ and\ \bibinfo {author}
  {\bibfnamefont{R.}~\bibnamefont{Grimm}},\ }%
  \bibfield{journal}{%
  \bibinfo {journal} {Phys. Rev. A}\ }%
  \textbf{\bibinfo {volume} {79}},\ \bibinfo {pages} {042718} (\bibinfo {year}
  {2009})%
  \bibAnnoteFile{NoStop}{Pilch:Obs}%
\bibitem{Bhaseen:Feshising}%
  \BibitemOpen
  \bibfield{author}{%
  \bibinfo {author} {\bibfnamefont{M.~J.}\ \bibnamefont{Bhaseen}}, \bibinfo
  {author} {\bibfnamefont{A.~O.}\ \bibnamefont{Silver}}, \bibinfo {author}
  {\bibfnamefont{M.}~\bibnamefont{Hohenadler}},\ and\ \bibinfo {author}
  {\bibfnamefont{B.~D.}\ \bibnamefont{Simons}},\ }%
  \bibfield{journal}{%
  \bibinfo {journal} {Phys. Rev. Lett.}\ }%
  \textbf{\bibinfo {volume} {103}},\ \bibinfo {pages} {265302} (\bibinfo {year}
  {2009})%
  \bibAnnoteFile{NoStop}{Bhaseen:Feshising}%
\bibitem{Zhou:PS}%
  \BibitemOpen
  \bibfield{author}{%
  \bibinfo {author} {\bibfnamefont{L.}~\bibnamefont{Zhou}}, \bibinfo {author}
  {\bibfnamefont{J.}~\bibnamefont{Qian}}, \bibinfo {author}
  {\bibfnamefont{H.}~\bibnamefont{Pu}}, \bibinfo {author}
  {\bibfnamefont{W.}~\bibnamefont{Zhang}},\ and\ \bibinfo {author}
  {\bibfnamefont{H.~Y.}\ \bibnamefont{Ling}},\ }%
  \bibfield{journal}{%
  \bibinfo {journal} {Phys. Rev. A}\ }%
  \textbf{\bibinfo {volume} {78}},\ \bibinfo {pages} {053612} (\bibinfo {year}
  {2008})%
  \bibAnnoteFile{NoStop}{Zhou:PS}%
\bibitem{Radzi:spinor}%
  \BibitemOpen
  \bibfield{author}{%
  \bibinfo {author} {\bibfnamefont{L.}~\bibnamefont{Radzihovsky}}\ and\
  \bibinfo {author} {\bibfnamefont{S.}~\bibnamefont{Choi}},\ }%
  \bibfield{journal}{%
  \bibinfo {journal} {Phys. Rev. Lett.}\ }%
  \textbf{\bibinfo {volume} {103}},\ \bibinfo {pages} {095302} (\bibinfo {year}
  {2009})%
  \bibAnnoteFile{NoStop}{Radzi:spinor}%
\bibitem{Timmermans:Feshbach}%
  \BibitemOpen
  \bibfield{author}{%
  \bibinfo {author} {\bibfnamefont{E.}~\bibnamefont{Timmermans}}, \bibinfo
  {author} {\bibfnamefont{P.}~\bibnamefont{Tommasini}}, \bibinfo {author}
  {\bibfnamefont{M.}~\bibnamefont{Hussein}},\ and\ \bibinfo {author}
  {\bibfnamefont{A.}~\bibnamefont{Kerman}},\ }%
  \bibfield{journal}{%
  \bibinfo {journal} {Phys. Rep.}\ }%
  \textbf{\bibinfo {volume} {315}},\ \bibinfo {pages} {199} (\bibinfo {year}
  {1999})%
  \bibAnnoteFile{NoStop}{Timmermans:Feshbach}%
\bibitem{Rad:Atmol}%
  \BibitemOpen
  \bibfield{author}{%
  \bibinfo {author} {\bibfnamefont{L.}~\bibnamefont{Radzihovsky}}, \bibinfo
  {author} {\bibfnamefont{J.}~\bibnamefont{Park}},\ and\ \bibinfo {author}
  {\bibfnamefont{P.~B.}\ \bibnamefont{Weichman}},\ }%
  \bibfield{journal}{%
  \bibinfo {journal} {Phys. Rev. Lett.}\ }%
  \textbf{\bibinfo {volume} {92}},\ \bibinfo {pages} {160402} (\bibinfo {year}
  {2004})%
  \bibAnnoteFile{NoStop}{Rad:Atmol}%
\bibitem{Romans:QPT}%
  \BibitemOpen
  \bibfield{author}{%
  \bibinfo {author} {\bibfnamefont{M.~W.~J.}\ \bibnamefont{Romans}}, \bibinfo
  {author} {\bibfnamefont{R.~A.}\ \bibnamefont{Duine}}, \bibinfo {author}
  {\bibfnamefont{S.}~\bibnamefont{Sachdev}},\ and\ \bibinfo {author}
  {\bibfnamefont{H.~T.~C.}\ \bibnamefont{Stoof}},\ }%
  \bibfield{journal}{%
  \bibinfo {journal} {Phys. Rev. Lett.}\ }%
  \textbf{\bibinfo {volume} {93}},\ \bibinfo {pages} {020405} (\bibinfo {year}
  {2004})%
  \bibAnnoteFile{NoStop}{Romans:QPT}%
\bibitem{Koetsier:Bosecross}%
  \BibitemOpen
  \bibfield{author}{%
  \bibinfo {author} {\bibfnamefont{A.}~\bibnamefont{Koetsier}}, \bibinfo
  {author} {\bibfnamefont{P.}~\bibnamefont{Massignan}}, \bibinfo {author}
  {\bibfnamefont{R.~A.}\ \bibnamefont{Duine}},\ and\ \bibinfo {author}
  {\bibfnamefont{H.~T.~C.}\ \bibnamefont{Stoof}},\ }%
  \bibfield{journal}{%
  \bibinfo {journal} {Phys. Rev. A}\ }%
  \textbf{\bibinfo {volume} {79}},\ \bibinfo {pages} {063609} (\bibinfo {year}
  {2009})%
  \bibAnnoteFile{NoStop}{Koetsier:Bosecross}%
\bibitem{Radzi:Resonant}%
  \BibitemOpen
  \bibfield{author}{%
  \bibinfo {author} {\bibfnamefont{L.}~\bibnamefont{Radzihovsky}}, \bibinfo
  {author} {\bibfnamefont{P.~B.}\ \bibnamefont{Weichman}},\ and\ \bibinfo
  {author} {\bibfnamefont{J.~I.}\ \bibnamefont{Park}},\ }%
  \bibfield{journal}{%
  \bibinfo {journal} {Ann. Phys.}\ }%
  \textbf{\bibinfo {volume} {323}},\ \bibinfo {pages} {2376} (\bibinfo {year}
  {2008})%
  \bibAnnoteFile{NoStop}{Radzi:Resonant}%
\bibitem{Dickerscheid:Feshbach}%
  \BibitemOpen
  \bibfield{author}{%
  \bibinfo {author} {\bibfnamefont{D.~B.~M.}\ \bibnamefont{Dickerscheid}},
  \bibinfo {author} {\bibfnamefont{U.}~\bibnamefont{{Al Khawaja}}}, \bibinfo
  {author} {\bibfnamefont{D.}~\bibnamefont{van Oosten}},\ and\ \bibinfo
  {author} {\bibfnamefont{H.~T.~C.}\ \bibnamefont{Stoof}},\ }%
  \bibfield{journal}{%
  \bibinfo {journal} {Phys. Rev. A}\ }%
  \textbf{\bibinfo {volume} {71}},\ \bibinfo {pages} {043604} (\bibinfo {year}
  {2005})%
  \bibAnnoteFile{NoStop}{Dickerscheid:Feshbach}%
\bibitem{Sengupta:Feshbach}%
  \BibitemOpen
  \bibfield{author}{%
  \bibinfo {author} {\bibfnamefont{K.}~\bibnamefont{Sengupta}}\ and\ \bibinfo
  {author} {\bibfnamefont{N.}~\bibnamefont{Dupuis}},\ }%
  \bibfield{journal}{%
  \bibinfo {journal} {Europhys. Lett.}\ }%
  \textbf{\bibinfo {volume} {70}},\ \bibinfo {pages} {586} (\bibinfo {year}
  {2005})%
  \bibAnnoteFile{NoStop}{Sengupta:Feshbach}%
\bibitem{Rousseau:Fesh}%
  \BibitemOpen
  \bibfield{author}{%
  \bibinfo {author} {\bibfnamefont{V.~G.}\ \bibnamefont{Rousseau}}\ and\
  \bibinfo {author} {\bibfnamefont{P.~J.~H.}\ \bibnamefont{Denteneer}},\ }%
  \bibfield{journal}{%
  \bibinfo {journal} {Phys. Rev. Lett}\ }%
  \textbf{\bibinfo {volume} {102}},\ \bibinfo {pages} {015301} (\bibinfo {year}
  {2009})%
  \bibAnnoteFile{NoStop}{Rousseau:Fesh}%
\bibitem{Rousseau:Mixtures}%
  \BibitemOpen
  \bibfield{author}{%
  \bibinfo {author} {\bibfnamefont{V.~G.}\ \bibnamefont{Rousseau}}\ and\
  \bibinfo {author} {\bibfnamefont{P.~J.~H.}\ \bibnamefont{Denteneer}},\ }%
  \bibfield{journal}{%
  \bibinfo {journal} {Phys. Rev. A}\ }%
  \textbf{\bibinfo {volume} {77}},\ \bibinfo {pages} {013609} (\bibinfo {year}
  {2008})%
  \bibAnnoteFile{NoStop}{Rousseau:Mixtures}%
\bibitem{Kuklov:Commens}%
  \BibitemOpen
  \bibfield{author}{%
  \bibinfo {author} {\bibfnamefont{A.}~\bibnamefont{Kuklov}}, \bibinfo {author}
  {\bibfnamefont{N.}~\bibnamefont{Prokof'ev}},\ and\ \bibinfo {author}
  {\bibfnamefont{B.}~\bibnamefont{Svistunov}},\ }%
  \bibfield{journal}{%
  \bibinfo {journal} {Phys. Rev. Lett.}\ }%
  \textbf{\bibinfo {volume} {92}},\ \bibinfo {pages} {050402} (\bibinfo {year}
  {2004})%
  \bibAnnoteFile{NoStop}{Kuklov:Commens}%
\bibitem{Messiah:QM}%
  \BibitemOpen
  \bibfield{author}{%
  \bibinfo {author} {\bibfnamefont{A.}~\bibnamefont{Messiah}},\ }%
  \emph{\bibinfo {title} {Quantum Mechanics}}\ (\bibinfo {publisher} {Dover},\
  \bibinfo {year} {1999})%
  \bibAnnoteFile{NoStop}{Messiah:QM}%
\bibitem{Chaikin:Book}%
  \BibitemOpen
  \bibfield{author}{%
  \bibinfo {author} {\bibfnamefont{P.~M.}\ \bibnamefont{Chaikin}}\ and\
  \bibinfo {author} {\bibfnamefont{T.~C.}\ \bibnamefont{Lubensky}},\ }%
  \emph{\bibinfo {title} {Principles of Condensed Matter Physics}}\ (\bibinfo
  {publisher} {Cambridge University Press},\ \bibinfo {year} {1995})%
  \bibAnnoteFile{NoStop}{Chaikin:Book}%
\bibitem{Auerbach:Interacting}%
  \BibitemOpen
  \bibfield{author}{%
  \bibinfo {author} {\bibfnamefont{A.}~\bibnamefont{Auerbach}},\ }%
  \emph{\bibinfo {title} {Interacting Electrons and Quantum Magnetism}}\
  (\bibinfo {publisher} {Springer},\ \bibinfo {year} {1994})%
  \bibAnnoteFile{NoStop}{Auerbach:Interacting}%
\bibitem{Duan:Control}%
  \BibitemOpen
  \bibfield{author}{%
  \bibinfo {author} {\bibfnamefont{L.~M.}\ \bibnamefont{Duan}}, \bibinfo
  {author} {\bibfnamefont{E.}~\bibnamefont{Demler}},\ and\ \bibinfo {author}
  {\bibfnamefont{M.~D.}\ \bibnamefont{Lukin}},\ }%
  \bibfield{journal}{%
  \bibinfo {journal} {Phys. Rev. Lett.}\ }%
  \textbf{\bibinfo {volume} {91}},\ \bibinfo {pages} {090402} (\bibinfo {year}
  {2003})%
  \bibAnnoteFile{NoStop}{Duan:Control}%
\bibitem{Lieb:Two}%
  \BibitemOpen
  \bibfield{author}{%
  \bibinfo {author} {\bibfnamefont{E.}~\bibnamefont{Lieb}}, \bibinfo {author}
  {\bibfnamefont{T.}~\bibnamefont{Schultz}},\ and\ \bibinfo {author}
  {\bibfnamefont{D.}~\bibnamefont{Mattis}},\ }%
  \bibfield{journal}{%
  \bibinfo {journal} {Ann. Phys.}\ }%
  \textbf{\bibinfo {volume} {16}},\ \bibinfo {pages} {407} (\bibinfo {year}
  {1961})%
  \bibAnnoteFile{NoStop}{Lieb:Two}%
\bibitem{Schultz:2D}%
  \BibitemOpen
  \bibfield{author}{%
  \bibinfo {author} {\bibfnamefont{T.~D.}\ \bibnamefont{Schultz}}, \bibinfo
  {author} {\bibfnamefont{D.~C.}\ \bibnamefont{Mattis}},\ and\ \bibinfo
  {author} {\bibfnamefont{E.~H.}\ \bibnamefont{Lieb}},\ }%
  \bibfield{journal}{%
  \bibinfo {journal} {Rev. Mod. Phys.}\ }%
  \textbf{\bibinfo {volume} {36}},\ \bibinfo {pages} {856} (\bibinfo {year}
  {1964})%
  \bibAnnoteFile{NoStop}{Schultz:2D}%
\bibitem{Pfeuty:TFI}%
  \BibitemOpen
  \bibfield{author}{%
  \bibinfo {author} {\bibfnamefont{P.}~\bibnamefont{Pfeuty}},\ }%
  \bibfield{journal}{%
  \bibinfo {journal} {Ann. Phys.}\ }%
  \textbf{\bibinfo {volume} {57}},\ \bibinfo {pages} {79} (\bibinfo {year}
  {1970})%
  \bibAnnoteFile{NoStop}{Pfeuty:TFI}%
\bibitem{Ovchinnikov:Mixed}%
  \BibitemOpen
  \bibfield{author}{%
  \bibinfo {author} {\bibfnamefont{A.~A.}\ \bibnamefont{Ovchinnikov}}, \bibinfo
  {author} {\bibfnamefont{D.~V.}\ \bibnamefont{Dmitriev}}, \bibinfo {author}
  {\bibfnamefont{V.~Y.}\ \bibnamefont{Krivnov}},\ and\ \bibinfo {author}
  {\bibfnamefont{V.~O.}\ \bibnamefont{Cheranovskii}},\ }%
  \bibfield{journal}{%
  \bibinfo {journal} {Phys. Rev. B}\ }%
  \textbf{\bibinfo {volume} {68}},\ \bibinfo {pages} {214406} (\bibinfo {year}
  {2003})%
  \bibAnnoteFile{NoStop}{Ovchinnikov:Mixed}%
\bibitem{Sondhi:Continuous}%
  \BibitemOpen
  \bibfield{author}{%
  \bibinfo {author} {\bibfnamefont{S.~L.}\ \bibnamefont{Sondhi}}, \bibinfo
  {author} {\bibfnamefont{S.~M.}\ \bibnamefont{Girvin}}, \bibinfo {author}
  {\bibfnamefont{J.~P.}\ \bibnamefont{Carini}},\ and\ \bibinfo {author}
  {\bibfnamefont{D.}~\bibnamefont{Shahar}},\ }%
  \bibfield{journal}{%
  \bibinfo {journal} {Rev. Mod. Phys.}\ }%
  \textbf{\bibinfo {volume} {69}},\ \bibinfo {pages} {315} (\bibinfo {year}
  {1997})%
  \bibAnnoteFile{NoStop}{Sondhi:Continuous}%
\bibitem{Sachdev:QPT}%
  \BibitemOpen
  \bibfield{author}{%
  \bibinfo {author} {\bibfnamefont{S.}~\bibnamefont{Sachdev}},\ }%
  \emph{\bibinfo {title} {Quantum Phase Transitions}}\ (\bibinfo {publisher}
  {Cambridge University Press},\ \bibinfo {year} {1999})%
  \bibAnnoteFile{NoStop}{Sachdev:QPT}%
\bibitem{Roth:Twocolor}%
  \BibitemOpen
  \bibfield{author}{%
  \bibinfo {author} {\bibfnamefont{R.}~\bibnamefont{Roth}}\ and\ \bibinfo
  {author} {\bibfnamefont{K.}~\bibnamefont{Burnett}},\ }%
  \bibfield{journal}{%
  \bibinfo {journal} {Phys. Rev. A}\ }%
  \textbf{\bibinfo {volume} {68}},\ \bibinfo {pages} {023604} (\bibinfo {year}
  {2003})%
  \bibAnnoteFile{NoStop}{Roth:Twocolor}%
\bibitem{Silver:BHC}%
  \BibitemOpen
  \bibfield{author}{%
  \bibinfo {author} {\bibfnamefont{A.~O.}\ \bibnamefont{Silver}}, \bibinfo
  {author} {\bibfnamefont{M.}~\bibnamefont{Hohenadler}}, \bibinfo {author}
  {\bibfnamefont{M.~J.}\ \bibnamefont{Bhaseen}},\ and\ \bibinfo {author}
  {\bibfnamefont{B.~D.}\ \bibnamefont{Simons}},\ }%
  \bibfield{journal}{%
  \bibinfo {journal} {Phys. Rev. A}\ }%
  \textbf{\bibinfo {volume} {81}},\ \bibinfo {pages} {023617} (\bibinfo {year}
  {2010})%
  \bibAnnoteFile{NoStop}{Silver:BHC}%
\bibitem{Fisher:Bosonloc}%
  \BibitemOpen
  \bibfield{author}{%
  \bibinfo {author} {\bibfnamefont{M.~P.~A.}\ \bibnamefont{Fisher}}, \bibinfo
  {author} {\bibfnamefont{P.~B.}\ \bibnamefont{Weichman}}, \bibinfo {author}
  {\bibfnamefont{G.}~\bibnamefont{Grinstein}},\ and\ \bibinfo {author}
  {\bibfnamefont{D.~S.}\ \bibnamefont{Fisher}},\ }%
  \bibfield{journal}{%
  \bibinfo {journal} {Phys. Rev. B}\ }%
  \textbf{\bibinfo {volume} {40}},\ \bibinfo {pages} {546} (\bibinfo {year}
  {1989})%
  \bibAnnoteFile{NoStop}{Fisher:Bosonloc}%
\bibitem{Kuhner:Phases}%
  \BibitemOpen
  \bibfield{author}{%
  \bibinfo {author} {\bibfnamefont{T.~D.}\ \bibnamefont{K\"uhner}}\ and\
  \bibinfo {author} {\bibfnamefont{H.}~\bibnamefont{Monien}},\ }%
  \bibfield{journal}{%
  \bibinfo {journal} {Phys. Rev. B}\ }%
  \textbf{\bibinfo {volume} {58}},\ \bibinfo {pages} {R14741} (\bibinfo {year}
  {1998})%
  \bibAnnoteFile{NoStop}{Kuhner:Phases}%
\bibitem{Kuhner:1DBH}%
  \BibitemOpen
  \bibfield{author}{%
  \bibinfo {author} {\bibfnamefont{T.~D.}\ \bibnamefont{K\"uhner}}, \bibinfo
  {author} {\bibfnamefont{S.~R.}\ \bibnamefont{White}},\ and\ \bibinfo {author}
  {\bibfnamefont{H.}~\bibnamefont{Monien}},\ }%
  \bibfield{journal}{%
  \bibinfo {journal} {Phys. Rev. B}\ }%
  \textbf{\bibinfo {volume} {61}},\ \bibinfo {pages} {12474} (\bibinfo {year}
  {2000})%
  \bibAnnoteFile{NoStop}{Kuhner:1DBH}%
\bibitem{Um:FiniteTFI}%
  \BibitemOpen
  \bibfield{author}{%
  \bibinfo {author} {\bibfnamefont{J.}~\bibnamefont{Um}}, \bibinfo {author}
  {\bibfnamefont{S.-I.}\ \bibnamefont{Lee}},\ and\ \bibinfo {author}
  {\bibfnamefont{B.~J.}\ \bibnamefont{Kim}},\ }%
  \bibfield{journal}{%
  \bibinfo {journal} {J. Korean Phys. Soc.}\ }%
  \textbf{\bibinfo {volume} {50}},\ \bibinfo {pages} {285} (\bibinfo {year}
  {2007})%
  \bibAnnoteFile{NoStop}{Um:FiniteTFI}%
\bibitem{Ejima:inprep}%
  \BibitemOpen
  \bibfield{author}{%
  \bibinfo {author} {\bibnamefont{{S. Ejima {\em et al}}}},\ }%
  \bibinfo {note} {in preparation}%
  \bibAnnoteFile{NoStop}{Ejima:inprep}%
\bibitem{Hieida:Reentrant}%
  \BibitemOpen
  \bibfield{author}{%
  \bibinfo {author} {\bibfnamefont{Y.}~\bibnamefont{Hieida}}, \bibinfo {author}
  {\bibfnamefont{K.}~\bibnamefont{Okunishi}},\ and\ \bibinfo {author}
  {\bibfnamefont{Y.}~\bibnamefont{Akutsu}},\ }%
  \bibfield{journal}{%
  \bibinfo {journal} {Phys. Rev. B}\ }%
  \textbf{\bibinfo {volume} {64}},\ \bibinfo {pages} {224422} (\bibinfo {year}
  {2001})%
  \bibAnnoteFile{NoStop}{Hieida:Reentrant}%
\bibitem{Blote:Conformal}%
  \BibitemOpen
  \bibfield{author}{%
  \bibinfo {author} {\bibfnamefont{H.~W.~J.}\ \bibnamefont{Bl{\"o}te}},
  \bibinfo {author} {\bibfnamefont{J.~L.}\ \bibnamefont{Cardy}},\ and\ \bibinfo
  {author} {\bibfnamefont{M.~P.}\ \bibnamefont{Nightingale}},\ }%
  \bibfield{journal}{%
  \bibinfo {journal} {Phys. Rev. Lett.}\ }%
  \textbf{\bibinfo {volume} {56}},\ \bibinfo {pages} {742} (\bibinfo {year}
  {1986})%
  \bibAnnoteFile{NoStop}{Blote:Conformal}%
\bibitem{Affleck:Universal}%
  \BibitemOpen
  \bibfield{author}{%
  \bibinfo {author} {\bibfnamefont{I.}~\bibnamefont{Affleck}},\ }%
  \bibfield{journal}{%
  \bibinfo {journal} {Phys. Rev. Lett.}\ }%
  \textbf{\bibinfo {volume} {56}},\ \bibinfo {pages} {746} (\bibinfo {year}
  {1986})%
  \bibAnnoteFile{NoStop}{Affleck:Universal}%
\bibitem{Cardy:scal}%
  \BibitemOpen
  \bibfield{author}{%
  \bibinfo {author} {\bibfnamefont{J.}~\bibnamefont{Cardy}},\ }%
  \emph{\bibinfo {title} {Scaling and Renormalization in Statistical Physics}}\
  (\bibinfo {publisher} {Cambridge University Press},\ \bibinfo {year} {1996})%
  \bibAnnoteFile{NoStop}{Cardy:scal}%
\bibitem{BPZ}%
  \BibitemOpen
  \bibfield{author}{%
  \bibinfo {author} {\bibfnamefont{A.~A.}\ \bibnamefont{Belavin}}, \bibinfo
  {author} {\bibfnamefont{A.~M.}\ \bibnamefont{Polyakov}},\ and\ \bibinfo
  {author} {\bibfnamefont{A.~B.}\ \bibnamefont{Zamolodchikov}},\ }%
  \bibfield{journal}{%
  \bibinfo {journal} {Nucl. Phys.}\ }%
  \textbf{\bibinfo {volume} {B241}},\ \bibinfo {pages} {333} (\bibinfo {year}
  {1984})%
  \bibAnnoteFile{NoStop}{BPZ}%
\bibitem{Francesco:CFT}%
  \BibitemOpen
  \bibfield{author}{%
  \bibinfo {author} {\bibfnamefont{P.~D.}\ \bibnamefont{Francesco}}, \bibinfo
  {author} {\bibfnamefont{P.}~\bibnamefont{Mathieu}},\ and\ \bibinfo {author}
  {\bibfnamefont{D.}~\bibnamefont{S{\'e}n{\'e}chal}},\ }%
  \emph{\bibinfo {title} {Conformal Field Theory}}\ (\bibinfo {publisher}
  {Springer},\ \bibinfo {year} {1997})%
  \bibAnnoteFile{NoStop}{Francesco:CFT}%
\bibitem{Muller:Sus}%
  \BibitemOpen
  \bibfield{author}{%
  \bibinfo {author} {\bibfnamefont{G.}~\bibnamefont{M\"uller}}\ and\ \bibinfo
  {author} {\bibfnamefont{R.~E.}\ \bibnamefont{Shrock}},\ }%
  \bibfield{journal}{%
  \bibinfo {journal} {Phys. Rev. B}\ }%
  \textbf{\bibinfo {volume} {30}},\ \bibinfo {pages} {5254} (\bibinfo {year}
  {1984})%
  \bibAnnoteFile{NoStop}{Muller:Sus}%
\bibitem{Lamacraft:OPS}%
  \BibitemOpen
  \bibfield{author}{%
  \bibinfo {author} {\bibfnamefont{A.}~\bibnamefont{Lamacraft}}\ and\ \bibinfo
  {author} {\bibfnamefont{P.}~\bibnamefont{Fendley}},\ }%
  \bibfield{journal}{%
  \bibinfo {journal} {Phys. Rev. Lett.}\ }%
  \textbf{\bibinfo {volume} {100}},\ \bibinfo {pages} {165706} (\bibinfo {year}
  {2008})%
  \bibAnnoteFile{NoStop}{Lamacraft:OPS}%
\bibitem{Rossini:Thermal}%
  \BibitemOpen
  \bibfield{author}{%
  \bibinfo {author} {\bibfnamefont{D.}~\bibnamefont{Rossini}}, \bibinfo
  {author} {\bibfnamefont{A.}~\bibnamefont{Silva}}, \bibinfo {author}
  {\bibfnamefont{G.}~\bibnamefont{Mussardo}},\ and\ \bibinfo {author}
  {\bibfnamefont{G.~E.}\ \bibnamefont{Santoro}},\ }%
  \bibfield{journal}{%
  \bibinfo {journal} {Phys. Rev. Lett.}\ }%
  \textbf{\bibinfo {volume} {102}},\ \bibinfo {pages} {127204} (\bibinfo {year}
  {2009})%
  \bibAnnoteFile{NoStop}{Rossini:Thermal}%
\bibitem{Calabrese:Quench}%
  \BibitemOpen
  \bibfield{author}{%
  \bibinfo {author} {\bibfnamefont{P.}~\bibnamefont{Calabrese}}\ and\ \bibinfo
  {author} {\bibfnamefont{J.}~\bibnamefont{Cardy}},\ }%
  \bibfield{journal}{%
  \bibinfo {journal} {Phys. Rev. Lett.}\ }%
  \textbf{\bibinfo {volume} {96}},\ \bibinfo {pages} {136801} (\bibinfo {year}
  {2006})%
  \bibAnnoteFile{NoStop}{Calabrese:Quench}%
\end{thebibliography}
\end{document}